\documentclass[12pt,preprint]{aastex}

\slugcomment{}

\shorttitle{Molecular Gas toward the SNR G347.3--0.5}
\shortauthors{Moriguchi et al.}

\begin{document}

\title{A Detailed Study of Molecular Clouds toward the TeV Gamma-Ray SNR G347.3--0.5}

\author{Y.\ Moriguchi, K.\ Tamura, Y.\ Tawara, H. Sasago, K. Yamaoka, T.\ Onishi, and Y.\ Fukui}
\affil{Department of Astrophysics, Nagoya University, Chikusa-ku, Nagoya 464-8602 \\ mori@a.phys.nagoya-u.ac.jp
}


\begin{abstract}

The supernovae remnant G347.3--0.5 (J1713.7--3946) is known as one of the unique SNRs which emit TeV gamma-ray as well as non-thermal X-rays.  We present a detailed study of molecular gas toward this SNR obtained with the 4m mm/sub-mm telescope NANTEN at an angular resolution of 2$\arcmin$.6.  This study has revealed that several intensity peaks and the overall distribution of the molecular gas with radial velocities from $-$12  km s$^{-1}$ to $-$3 km s$^{-1}$ show a remarkably good correlation with the X-ray features, strongly supporting the kinematic distance around 1 kpc derived by Fukui et al. (2003), as opposed to 6 kpc previously claimed. In addition, we show that absorption of X-rays is caused by local molecular gas at softer X-ray bands.  Subsequent measurements of the sub-mm $J$ = 3--2 transition of CO made with the ASTE 10 m and CSO 10.4 m sub-mm telescopes toward three of the molecular intensity peaks have revealed higher excitation conditions, most likely higher temperatures above $\sim$ 30 K as compared to that of the typical gas temperature, 10 K, in low mass dark clouds.  This temperature rise is most likely caused by enhanced heating by the high energy events in the SNR, where possible mechanisms include heating by X-rays, gamma-rays, and/or cosmic ray protons, although we admit additional radiative heating by young protostars embedded may be working as well.

In one of the CO peaks, we have confirmed the presence of broad molecular wings of $\sim$ 20 km s$^{-1}$ velocity extent in the CO $J$=3--2 transition.  Two alternative interpretations for the wings are presented; one is shock acceleration by the blast wave and the other is molecular outflow driven by an embedded protostar.  The SNR evolution is well explained as the free expansion phase based on the distance of 1 kpc.  The molecular dataset should be valuable to make a further detailed comparison with the gamma-ray and X-ray distributions in order to examine the cosmic ray acceleration quantitatively.

\end{abstract}



\section{Introduction}

Supernovae are generally thought to be the most energetic events in the Galactic disk and supernova remnants (SNRs) have profound effects on the dynamics and physical/chemical processes in the interstellar space through direct interactions with the surroundings, providing a unique laboratory to test high energy processes related to strong shocks.   Because of the large energy release of 10$^{51}$ erg, SNRs have also been considered as the origin of Galactic cosmic rays (e.g.,  Shklovsky 1953; Hayakawa  1956; Ginzburg 1957).  Cosmic rays themselves have brought crucial information on elementary particles and in addition play a crucial role in the evolution of the interstellar medium mainly via heating and ionization, although detailed physics of particle acceleration in SNRs remains unclear.  It is therefore one of the key issues in astrophysics to elucidate the interaction of SNRs with their ambient matter and the probable production of cosmic rays therein.

Observations of the interstellar gas have revealed the interactions between SNRs and the surroundings under various physical conditions including ionized hot gas, neutral atomic gas and dense molecular gas (e.g., Rho et al. 1994 (optical and X-ray), Koo \& Heiles 1991 (H{\small \,I}),van Dishoeck et al. 1993 (molecular line)).
In particular, millimeter and sub-mm wave observations of dense molecular gas have proved to be a powerful probe of the shocked gas and the interactions; broad wings of several 10 km s$^{-1}$ accelerated by the shocks are observed in several SNRs in the mm and sub-mm spectra of interstellar CO and other molecules (e.g., W44; Seta et al. 1998, W28; Arikawa et al. 1999, IC443; White et al. 1987).  It is also to be noted that observations of molecular gas are extremely useful in constraining the kinematic distance of SNRs owing to the better angular resolutions of mm-wave telescopes and the intrinsic smaller velocity dispersions of the molecular gas compared to atomic gas.

On the other hand, observational indications for the acceleration of high-energy particles in SNRs have been quite poor particularly for cosmic ray protons.  Cosmic rays consist of protons as the major constituent as well as of the other minor constituents including electrons and heavier atomic nuclei.   For years, observational indications for the cosmic ray particles in SNRs were available only for the electrons that emit synchrotron radiation in the radio band.   The highest energy of such electrons are however in the order of MeV, much below the highest energy of the Galactic cosmic rays around 10$^{15}$ eV.  The detection of nonthermal X-ray emission from the shell-type SNR SN 1006 provided evidence for such very high energy electrons (e.g., Koyama et al.  1995).  Considering the very short lifetime of such high energy electrons the site of their prediction is definitely constrained in the SNR itself.   We are however still lacking of understanding of the origin of the major constituent, cosmic ray protons, as there is not yet clear evidence for proton acceleration in SNRs.

Two shell-type remnants with nonthermal-dominant X-ray spectra have been identified subsequently; they are G347.3--0.5 (Koyama et al.  1997; Slane et al.  1999) and G266.2--1.2 (Slane et al. 2001).   G347.3--0.5 (RX J1713.7--3946) was first discovered in the ROSAT All-Sky Survey by Pfeffermann \& Aschenbach (1996).   Later, ASCA observations revealed that the X-ray emission from the remnant is predominantly nonthermal (Koyama et al. 1997; Slane et al. 1999).   The remnant is $\sim$ 1$^{\circ}$ in diameter and appears to be of a shell-type morphology with the brightest emission in the western region (We shall use the equatotial coordinates to identify directions in the sky throughout the present paper).  Uchiyama et al. (2003) showed that the synchrotron cutoff energy is unusually high beyond 10 keV, corresponding to 10$^{14}$ eV as the highest particle energies.   G347.3--0.5 was also detected at the TeV $\gamma$-ray range with the CANGAROO telescope (Muraishi et al. 2000; Enomoto et al. 2002), and models of broadband emission point to IC scattering as the origin of the TeV $\gamma$-ray photons (Muraishi et al. 2000; Ellison et al. 2001).   More recently, follow-up TeV $\gamma$-ray observations have led to a different conclusion, suggesting pion decay as the source of energetic photons (Enomoto et al. 2002), but the nature of this emission is still uncertain (see Butt et al. 2002; Reimer \& Pohl 2002; Aharonian et al. 2004).

The SNR is located at l $\sim$ 347$\arcdeg$, towards the direction of the central part of the Galaxy, a very crowded region with various galactic objects in the line of sight.  This makes it particularly confusing and uncertain to identify physically associated objects and accordingly, the distance was rather poorly determined at best.  The determination of the distance is crucial in all the interpretation of observations.  Originally, Pfeffermann \& Aschenbach (1996) adopted a distance to the SNR of $\sim$ 1 kpc based on the estimation of the column density derived from the spectral analysis of the ASCA data (see also Koyama et al. 1997).  Later, the possible association of a molecular cloud (later refered to as "cloud A") was used to derive a distance to be 6.3 $\pm$ 0.4 kpc from observations of the 2.6 mm CO($J$=1--0) line emission at 9 arc min resolution by Slane et al. (1999).  In their discussion, these authors argued that an enhanced value of CO($J$=2--1)/($J$=1--0) line ratio in cloud A indicates the physical interaction with the SNR shock (see also Butt et al. 2001).  This estimation, $\sim$ 6 kpc, was often used in the following studies of the physical properties of G347.3--0.5 (e.g., Lazendic et al. 2004; Pannuti et al. 2003, Cassam-Chenai et al. 2004a).

Most recently, Fukui et al. (2003, hereafter Paper I) performed new CO observations at 2.6 mm wavelength with NANTEN, a 4-meter mm and sub-mm telescope located at Las Campanas Observatory in Chile.  These new observations have a resolution of 2$\arcmin$.6, a factor of 3.4 better than that of the CO data used by Slane et al. (1999).   The authors discovered molecular gas at a distance of $\sim$ 1 kpc which shows hole-like distribution having a striking correlation with the X-ray image of G347.3--0.5, in addition to the presence of a broad CO line in one of the CO peaks that likely represents dynamical interaction between the SNR shock and the ambient molecular gas.  They also find that cloud A shows worse spatial contact with the edge of the SNR than in the lower resolution CO data.   In addition, Koo et al. (2003), Koo et al. (2004) and Cassam-Chenai et al. (2004b) derive a similar distance $\sim$ 1 kpc by analysing  and comparing it with X-ray absorption.   Most recently, the H.E.S.S. collaboration made a better resolution image of TeV $\gamma$-rays and has supported the interaction of the SNR with the CO clouds at $\sim$ 1 kpc (Aharonian et al. 2004).

Subsequent to Paper I, we present in this paper a detailed analysis of the CO distributions and the physical properties of molecular clouds associated with G347.3--0.5.  We present the observations in section 2, and the results of the observations in section 3.   We discuss the distance, the evolutionary phase, and the surrounding environments of the SNR in section 4 and give conclusions in section 5.


\section{Observations}

Observations were made in the $^{12}$CO ($J$=1--0) line by using a 4-meter mm/sub-mm telescope of Nagoya University, NANTEN, at Las Campanas Observatory, Chile.   The telescope had a half-power beam width (HPBW) of 2.$^{\prime}$6 at a frequency of 115 GHz and was equipped with a 4 K cryogenically cooled Nb superconductor-insulator-superconductor (SIS) mixer receiver (Ogawa et
al. 1990) that provided a typical system temperature of $\sim$ 250 K in the single-side band, including the atmosphere toward the zenith.  NANTEN was equipped with two acousto-optical spectrometers (AOS) with 2048 channels.  The total bandwidth and the effective resolution were 250 MHz and 250 kHz for a wide-band mode, corresponding to a velocity coverage of 600 km s$^{-1}$ and a velocity resolution of 0.65 km s$^{-1}$, respectively.  Present observations used the wide-band mode.
The data is part of the NANTEN Galactic Plane Survey.  The survey carried out from 1999 to 2003 observed about 1.1 million points in total, covering a galactic longitude range of 240 degrees (from $l$ = 180$^{\circ}$ to 60$^{\circ}$) and galactic latitude coverage of 10--20 degrees at 4$\arcmin$--8$\arcmin$ (partly 2$\arcmin$) grid-spacing.  The grid spacing for the longitude coverage is basically 4$^{\prime}$ for $|b| \leq$ 5$^{\circ}$, 8$^{\prime}$ for $|b| \geq$ 5$^{\circ}$, respectively. (see Matsunaga et al. and the other papers in PASJ Vol. 53 No.6 2001)

After the 4$\arcmin$ grid unbiased survey, we carried out a sensitive 2$\arcmin$ grid survey, by covering the whole area of G347.3--0.5 in April 2003.   An area of $\sim$ 1.91 square degrees in the region of 346$^{\circ}.7 \leq l \leq$ 348$^{\circ}$.0 and $-$1$^{\circ}.2 \leq b \leq$ 0$^{\circ}$.2 were mapped.  The integration time per point was typically $\sim$ 10 s/point, resulting in a typical rms noise fluctuations of 0.2--0.3 K at a velocity resolution of 0.65 km s$^{-1}$.  In total, 1720 positions were observed.  For the intensity calibration, the room-temperature chopper wheel method was employed and the absolute intensity calibration was made by observing Orion KL [$\alpha$(1950) = 5$^{\rm h}$32$^{\rm m}$47.$^{\rm s}$0, $\delta$(1950) = $-$5$^{\circ}$24$^{\prime}$21$^{\prime\prime}$] and $\rho$ Oph East [$\alpha$(1950) = 16$^{\rm h}$29$^{\rm m}$20$^{\rm s}$.9,  $\delta$(1950) = $-$24$^{\circ}$22$^{\prime}$13$^{\prime\prime}$] every a few hours.  We assumed the absolute radiation temperatures, $T_{\rm R}^*$, of Orion KL and $\rho$ Oph East to be 65 K and 15 K, respectively.

Following the NANTEN survey, we have carried out a $^{12}$CO($J$=3--2) one-point observation toward one of the strong $^{12}$CO($J$=1--0) peaks identified with NANTEN, peak C discussed in Paper I, using the 10.4m telescope at the Caltech Submilimeter Observatory (CSO) in April 2004.  CSO is installed at the summit of Mt. Mauna Kea at an altitude of 4,000m.  The observed position is ($l$, $b$) = (347$\arcdeg$.07, $-$0$\arcdeg$.400) and the beam size at the $^{12}$CO($J$=3--2) line was 22$\arcsec$.  CSO was equipped with four acousto-optical spectrometers (AOS) with differenet bandwidths, 50 MHz, 500 MHz (two), and 1.5 GHz, respectively.  The number of the channels is 2048 for 1.5 GHz and 500MHz, 1024 for 50 MHz, respectively.  We used two of these AOS's, 1.5GHz and 50MHz.  The effective ferequency resolution were 700 kHz for the 1.5GHz AOS and 48 kHz for the 50MHz one, respectively.  The velocity coverage and resolution were 1200 km s$^{-1}$ and 0.61 km s$^{-1}$ for 1.5GHz, 40 km s$^{-1}$ and 0.04 km s$^{-1}$ for 50 MHz, respectively.  The system temperature for the 1.5 GHz was $\sim$ 1200 K for an elevation angle of $\sim$ 26$\arcdeg$ in the double-side band including the atmosphere.  The total integration time is 120 s and the rms noise temperature per channel is 0.25 K for 1.5GHz and 0.55 K for 50 MHz, respectively.  We adopted a beam efficiency $\sim$ 75\% for 345 GHz, the recommended value calculated by past observations of planets. 

Successively, we have carried out high resolution-extensive $^{12}$CO($J$=3--2)  survey toward G347.3--0.5 using ASTE 10-meter submillimeter telescope of NAOJ (Ezawa et al. 2004) located at the site of Atacama desert in Chile, with an altitude of 4800m, in November 2004.  The beam size at the $^{12}$CO($J$=3--2) line was 23$\arcsec$.  We have observed the three regions containing the CO peaks selected from Paper I with a grid spacing of 30$\arcsec$ (partially 1$\arcmin$), including 780 observed points in total.  ASTE was equipped with four digital backend system (auto-correlator) with 2048 channels and we could select either of the two observational modes (wide-band or narrow-band).  The total bandwidths of the wide-band mode and the narrow-band one were 512 MHz and 128 MHz corresponding to an effective velocity coverages of 450 km s$^{-1}$ and 110 km s$^{-1}$, respectively.  The effective resolutions were 500 kHz and 125 kHz, corresponding to a velocity resolution of 0.43 km s$^{-1}$ and 0.12 km s$^{-1}$, respectively.  We used the wide-band mode.  The system temperature was 300-400 K for an elevation angle of $\sim$ 40-70$\arcdeg$ in the double-side band including the atmosphere.  Total integration time per point is 30 s and the rms noise temperature per channel is 0.4-0.9 K, respectively.  The beam efficiency was $\sim$ 50-60\% for 345 GHz in daytime.  For system check and absolute intensity calibration, M17 SW [$\alpha$(1950) = 18$^{\rm h}$17$^{\rm m}$30.$^{\rm s}$0, $\delta$(1950) = $-$16$^{\circ}$13$^{\prime}$6$^{\prime\prime}$] was observed every day.


\section{Results}

\subsection{Large Scale CO Distribution}

Figure 1(a) shows a CO map covering from $l$ = 343$\arcdeg$ to $l$ = 352$\arcdeg$ taken with NANTEN as part of the galactic plane CO survey (Fukui et al. 2005, in preparation) superposed on the ROSAT X-ray boundary of SNR G347.3--0.5.  The SNR happens to be located toward a hole of the integrated CO intensity.
Figure 1(b) indicates a velocity-longitude diagram in the same longitude range with Figure 1(a) for a velocity ( = radial velocity with respect to the Local Standard of Rest = $V_{\rm LSR}$) range from -150 km s$^{-1}$ to 20 km s$^{-1}$.  The velocity range of the molecular gas argued to be interacting with the SNR is close to $\sim$ 0 km s$^{-1}$, from $-$12 to $-$3 km s$^{-1}$ (Paper I).   It is seen that the lower negative velocity edge at $V_{\rm LSR}$ $\sim$ from 0 km s$^{-1}$ to $-$10 km s$^{-1}$ of the Sagittarius arm exhibits an intensity depression in $l$ $\sim$ 347$\arcdeg$-348$\arcdeg$.  This corresponds to a cavity-like CO distribution delineating the SNR G347.3--0.5 in Paper I, having a size similar to that of the SNR.   
More significant is a hole of CO emission at $V_{\rm LSR}$ from $\sim$ $-$10 km s$^{-1}$ to $\sim$ $-$35 km s$^{-1}$ in the same longitude range of l $\sim$ 347$\arcdeg$-348$\arcdeg$, associated with the Sagittarius arm.  This hole corresponds to a molecular supershell named as SG 347.3--0.0--21 at a distance of $\sim$ 3 kpc (Matsunaga et al. 2001), which is likely due to multiple supernovae/stellar winds by OB stars.   This chance coincidence with the supershell in the line of sight causes the apparent CO depression toward G347.3--0.5 in Figure 1(a), while the size of the supershell, $\sim$ 1.5$\arcdeg$, is much larger than that of G347.3--0.5 and they are not likely physically connected.
Another feature of past concern is cloud A located at $l$ $\sim$ 348$\arcdeg$ and $V_{\rm LSR}$ $\sim$ $-$90 km s$^{-1}$, which was ascribed to be associated with the SNR G347.3--0.5 by Slane et al. (1999).   This cloud is rather isolated on the far side of the Norma arm.

\subsection{CO Clouds towards the SNR: Coarse Velocity Channel Maps}

In Figure 2, we present a montage of CO velocity channel maps taken every 10 km s$^{-1}$ from $V_{\rm LSR}$ =-160 km s$^{-1}$ to 20 km s$^{-1}$.  The directions seen from the center of the SNR in the equatrical coordinates are denoted in Figure 2(a), such as N (north), NE (northeast), and so on. The central position is denoted by the cross, where we adopt ($l$, $b$) = (347.3, $-$0.5), the catalogued position originally given in Pfeffermann \& Aschenbach (1996). 
  
The kinematic distance corresponding to the velocity centroid at each panel is indicated as calculated by using the Galactic rotation curve model (Brand \& Blitz 1993).  
The outer boundary of the ROSAT X-ray image is superposed in each panel for reference.   At various velocities, the SNR appears to be in contact with CO features as naturally expected in the direction of the galactic center, but most of the apparent coincidence must be fortuitous.

At $V_{\rm LSR}$ $\sim$ $-$160 to $-$110 km s$^{-1}$ (a)--(e) the association of CO with the SNR is not obvious.  At $-$100 to $-$90 km s$^{-1}$ (g), a CO cloud towards ($l$, $b$) = (347$\arcdeg$.9, $-$0$\arcdeg$.3) appears close to the boundary of the SNR, which is cloud A.  The present data taken at a higher angular resolution than that of Slane et al. (1999) shows that the western side of cloud A is not well delineating the SNR boundary compared to the lower resolution data at 9$\arcmin$ particularly an a latitude range greater than $-$0$\arcdeg$.3.  
Another CO feature is seen at $V_{\rm LSR}$ $\sim$ $-$80 to $-$70 km s$^{-1}$ (i) towards ($l$, $b$) = (347$\arcdeg$.30, 0$\arcdeg$.05), which was also suggested to be interacting with the SNR (Slane et al. 1999), although the kinematic distances of cloud A and the $-$70 km s$^{-1}$ cloud differ by $\sim$ 3 kpc, making it highly unlikely that both of the clouds are a connected single cloud interacting with the SNR.  
At $V_{\rm LSR}$ $\sim$ $-$60 to $-$50 km s$^{-1}$ (k), weak CO features appear to surround the SNR on the northwest, while this was not noted before.  The panel at $V_{\rm LSR}$ $\sim$ $-$50 to $-$40 km s$^{-1}$ (l) shows little CO emission and these velocities correspond to the inter arm.  At $V_{\rm LSR}$ $\sim$ $-$40 to $-$20 km s$^{-1}$ (m and n) the supershell SG 347.3--0.5--21 mentioned above are seen.  This supershell has a significantly larger radius than that of the SNR G347.3--0.5 while some of the CO features appear to overlap with the SNR perhaps by chance. 
What is most remarkable a panel (p) at $-$10 to $-$0 km s$^{-1}$ which shows that the CO distribution appears to well delineate the SNR boundary.  For over three quarters of the SNR boundary except for the southeast edge, the X-ray image seems to be bound by the CO emission.  This is the velocity component ascribed to be interacting with the SNR in Paper I.   More details of the features will be shown in the following sections (section 3.3 and 3.4).  We also note a strong CO peak at ($l$, $b$) = (347$\arcdeg$.1, $-$0$\arcdeg$.4) in panel (o) is likely associated with the SNR (see section 3.4).
Beyond $V_{\rm LSR}$ $\sim$ 0 km s$^{-1}$, the CO emission shows a fairly good correspondence with the depression of the X-ray image towards $l$ $\sim$ 347$\arcdeg$.0 to 347$\arcdeg$.4 and $b$ $\sim$ $-$0.$\arcdeg$6 to $-$0$\arcdeg$.7, which is elongated by half a degree from the southwest to east.  This depression is ascribed to the absorption by a local molecular gas at 0 to 10 km s$^{-1}$ in section 3.6.  
We note that most of the CO emission seen in panel (q) is due to the local clouds independent of the others at minus velocities.  The local clouds have narrow linewidths of $\sim$ 1.5 km s$^{-1}$ and peak velocity of $\sim$ 6 to 7 km s$^{-1}$, clearly discerned from the distant clouds in $V_{\rm LSR}$ $<$ 0 km s$^{-1}$, as seen in Figure 1(b).  

Figure 1(a) shows that the molecular distribution towards G347.3--0.5 generally shows a depression toward the SNR and Figure 2 indicated the same trend.  We shall here introduce a factor to describe the degree of geometrical contact between the CO and X-ray images, which may help one to test how the molecular gas is in contact with the SNR at each distance.  This factor, a "covering factor", is defined as a ratio of a covering angle of the CO distribution surrounding the SNR boundary to the whole angle of the SNR in each velocity range every 10 km s$^{-1}$.  We adopt the center of the angular measurement to be ($l$, $b$) = (347$\arcdeg$.3, $-$0$\arcdeg$.5).  For simplicity, this covering angle is calculated only for the overlapping region of the SNR boundary and CO lowest contours in the channel maps shown in Figure 2.  The contributions from the isolated clouds located inside the X-ray boundary are not included. Figure 3(a)$-$(b) are schematics for estimating the covering factor.  The factor is $\sim$ 10\% in the case of Figure 3(a) and 80\% in Figure 3(b), respectively. 

Since the distribution of molecular clouds toward G347.3--0.5 generally tends to surround the SNR and to avoid the interior region of the SNR in any velocity range, except for the positive velocity corresponding to the local cloud components, the covering factor gives a quantitative measure of the overall spatial coincidence between CO and X-ray around the boundary of the SNR.   
The covering factor is shown as a function of $V_{\rm LSR}$ in Figure 4.  For most velocity ranges, the factor is less than $\sim$ 40\%, but at around $-$10 km s$^{-1}$ it becomes largest, $\sim$ 80\%, indicating that the correlation of CO with the SNR is the highest at that velocity range in panel (p) of Figure 2.   We suggest this lends another support for the assignment of the low velocity CO emission as interacting with the SNR in Paper I.

\subsection{Detailed Comparisons with X-rays}

In order to make a more detailed examination of the spatial correlation between the X-ray features and the interacting molecular gas at velocity range from $\sim$ $-$20 km s$^{-1}$ to $\sim$ 0 km s$^{-1}$, we show in Figure 5 a superposition of ROSAT image for the whole SNR and CO and in Figure 6 a superposition of the XMM image at harder energy range (2-7 keV) and at higher angular resolution (15$\arcsec$) (Hiraga et al. 2004).  Figure 5(a) shows the gray scale and contour map of $^{12}$CO($J$=1--0) integrated intensity and Figure 5(b) shows a superposition of the ROSAT X-ray image at an energy range of 0.1-2.0 keV on the CO for a velocity range from $-$12 km s$^{-1}$ to $-$3 km s$^{-1}$ which as we find shows the best correlation with the SNR. 

In order to estimate the physical parameters of the molecular gas interacting with the SNR, we need to identify molecular clouds.  As shown in Figures 5 and 6, we named CO peaks from A through Y, while we admit the complicated CO distribution makes it difficult to identify clouds uniquely.  A, B, C, and D, have been used in Paper I, and in addition to these, we indicate other peaks of weaker intensity as E through Y.  Basic physical properties of the clouds tentatively identified are listed in Table 1.  We note the present cloud definition has not signifacant influence on later discussion of the physical aspects of the interaction.  

The method of defining a CO cloud is as follows:

(1) In Figure 5, an area enclosed by the contour of 8.5 K km s$^{-1}$ (equal to 5 $\sigma$ level for the integration for a velocity range of 8 km s$^{-1}$), which is significantly above the extended emission, is defined as an individual cloud.

(2) In case that a CO peak has relatively weak intensity ($\leq$ 6.5 K km s$^{-1}$), the area enclosed by the contour of 4.5 K km s$^{-1}$ is identified as an individual cloud (O, Q, R, S, T, U, and V). 

(3) For a cloud for which it is hard to define a closed boundary with the contour of 8.5 K (because the area enclosed by cotour is too small or extended filamentarily), a circular area with a radius of 4$\arcmin$, slightly larger than the beam size, around the peak position is defined as a size of the cloud for simplicity (F, I, H, K1, and K2).

(4) When a peak position is separated by 0.3$\arcdeg$ or larger from the nearest boundary of the ROSAT X-ray image in Figure 5(b), the cloud is regarded as not associated directly with the SNR and is not taken into account (e.g., ($l$, $b$) $\sim$ (346$\arcdeg$.73, 0$\arcdeg$.02), (347$\arcdeg$.71, 0$\arcdeg$.10)).

(5) When a CO spectrum has a double-peaked profile, each velocity component is defined as an independent cloud in case that the velocity difference between the two peaks is larger than 5 km s$^{-1}$, $\sim$ 2.5 times larger than the typical linewidth of a cloud, (K1, K2, N1, and N2).  The two components are devided at the velocity of the intensity minumum between the peaks.

(6) When two or more peaks are contained in one area enclosed by a contour, the area devided at the minimum in the intensity map are defined as independent peaks. 

In Figure 5, The stronger CO peaks tend to be located close to the X-ray intensity peaks on the west and northwest.  The most prominent X-ray peaks at ($l$, $b$) = (347$\arcdeg$.1, $-$0$\arcdeg$.3) and (347$\arcdeg$.3, $-$0$\arcdeg$.05) are well correlated with CO peaks; X-ray peak (347$\arcdeg$.1, $-$0$\arcdeg$.3) is associated with peaks A and C.  B is not obvious here but as shown later B is associated with an X-ray peak at harder energy band.  The X-ray peak at (347$\arcdeg$.3, $-$0$\arcdeg$.05) is associated with D and L, where D is coincident with the CANGAROO and H.E.S.S. TeV $\gamma$-ray peak (Paper I, Aharonian et al. 2004).  Most of these CO peaks including E and G are located just on the outer edge of the SNR, as is consistent with a picture that the SNR shock is impacting these molecular clumps from the inward.  We also note that peaks O, Q and R correspond to the X-ray peaks at ($l$, $b$) = (347$\arcdeg$.52, $-$0$\arcdeg$.30) , (347$\arcdeg$.72, $-$0$\arcdeg$.50), and (347$\arcdeg$.60, $-$0$\arcdeg$.73), respectively.   On the south, the molecular gas including CO peaks S-Y also well delineates the SNR boundary.  To summarize, at scales of arc min, the X-ray image shows a good spatial correlation with NANTEN CO image.

Figure 6 shows an even more striking correspondence between CO and X-ray at 10$\arcsec$ scales taken with XMM at the energy range harder than that in Figure 5.  This X-ray image should be less affected by foreground absorption than the ROSAT image.  The CO peaks A and B are well delineated by the thin X-ray filament, which becomes clear in this harder energy range, extending by 15$\arcmin$ in the northwest-southeast.  This thin filament is not well traced in Figure 5 perhaps due to the absorption effect in the southeastern half of the filament facing CO peak B and also in part due to lower resolution of ROSAT.  Peak C appears surrounded by two thin X-ray features of a few to several arcmin lengths on the north-west and south-east.   Peak D is also X-ray bright on the south.  The other peaks are all located just outside the X-ray features.   The inside  of the SNR around the point source (1WGA J1713.4--3949) are almost empty in the CO distribution.  To summarize, this comparison further confirms the remarkable spatial correlation seen in Figure 5 at a finer angular resolution by a factor of three.

\subsection{Sub-mm Results; CO $J$=3--2 Emission}

Three regions of probable sites of the interaction, $^{12}$CO($J$=1--0) peaks of A, C, and D, have been mapped in the $^{12}$CO($J$=3--2) emission with ASTE.  The same transition has also been observed toward peak C with the CSO 10.4m telescope.  The intensity distributions of the CO ($J$=3--2) emission toward peak A, C, and D are shown in Figure 7.  The ratios of the integrated intensities of the $J$=3--2 and $J$=1--0 emission are $\sim$ 0.7, $\sim$ 0.8, and $\sim$ 0.5, towards peaks A, C, and D, respectively, where each of the CO intensity has been integrated within a contour at half-intensity levels (11 $\sigma$, 8 $\sigma$, and 6 $\sigma$ level in the $J$ = 3--2 distribution, respectively.)  We also note that the $J$=3--2 line is generally not excited to be observable with the current typical sensitivity in most of the cold molecular clouds in the Galaxy; Towards peak C, the $J$ = 3--2 emission is seeen at $V_{\rm LSR}$ = $-$11 km s$^{-1}$ and $-$90 km s$^{-1}$, only toward two of the four components seen in the $J$ = 1--0 emission (Figure 8).

Here we shall use the large velocity gradient (LVG) radiative transfer model to obtain constraints on the physical conditions in the molecular gas. 
The LVG model is a useful method to solve the equation of radiation and collision for molecular excitation states.  In the LVG model, it is assumed that the molecular cloud has the velocity gradient, and that all photons radiated in the cloud escape without absorption because a different position has a different Doppler velocity shift.  In this analysis, we use a LVG calculation method developed by Kim et al. (2002) based on Goldreich \& Kwan (1974).  
In the model, the radiative transfer equation gives population of each excitation state as a function of the kinetic temperature of the gas, $T_{\rm kin}$, for the molecular hydrogen number density $n({\rm H_{2}})$.  For each observed point of CO ($J$=3--2) and ($J$=1--0), we invert these functions to determine the  $T_{\rm kin}$ and  $n({\rm H_{2}})$ corresponding to the observed line ratios (see Figure 9).  The model assumes a plane-parallel cloud geometry.  The parameter needed for the model is the ratio $X$(CO)/$V$, where $X$(CO) is the fractional CO abundance and $V$ is the velocity gradient.  We adopt $X$(CO)/$V$ = 10$^{-4.5}$ pc km$^{-1}$ s in this case.

With the LVG analysis, we find that peaks A and C require rather high density and temperature as listed in Table 2.  The lowest temperature for peak A must exceed 30 K if we adopt typical density less than $\sim$ 10$^{3}$ cm$^{-3}$; we note that the typical density and temperature in local $^{12}$CO emitting dark clouds are 10$^{3}$ cm$^{-3}$ and 10 K, respectively (e.g., Snell et al. 1981).   
This indicates that the high-$J$ transition of CO is appreciably excited in the peaks A and C, and that the gas is warm, being consistent with the shocked molecular gas (e.g., Arikawa et al. 1999, Irwin \& Avery 1992, Zhu et al. 2003).  In particular, peak A showing the good correlation with the X-ray bright filament requires higher temperature than in typical quiescent molecular gas in the solar vicinity. 
In the excitation condition of typical cold molecular gas in nearby dark clouds with low mass star formation ($T_{\rm kin}$ $\sim$ 10-20 K, $n(\rm H_{2})$ $\sim$ 10$^{3-4}$ cm$^{-3}$ ), the $J$=3--2 transition is only weakly excited and the line intensity ratio between the $J$=3--2 and $J$=1--0 lines is likely below 0.3, significantly less than what is observed in the three peaks of G347.3--0.5 (see Figure 9).  Unfortunately, the observational data in the CO($J$=3--2) emission is quite limited over the Galaxy and only a limited number of areas have been observed so far (e.g., Hunter et al. 1997; Yamaguchi et al. 2003).  In order to make an estimate on the general CO $J$=3--2/$J$=1--0 ratio in typical molecular clouds in the solar vicinity which are not subject to extreme excitation conditions, we shall here integrate the CO emission of the present $J$=3--2 data and NANTEN $J$=1--0 data of the three regions toward G347.3--0.5 (shown in Figure 7) by eliminating the regions of probable interaction in velocity space.  The results are as follows; the $J$=3--2/$J$=1--0 ratios are $<$ 4.2 K km s$^{-1}$/130 K km s$^{-1}$ = 0.03 in the local clouds in $V_{\rm LSR}$ = 0-10 km s$^{-1}$ (4.2 K km s$^{-1}$ corresponds to a 3 $\sigma$ detection limit of $J$=3--2 emission over the integrated velocity range), and 23 K km s$^{-1}$/91 K km s$^{-1}$ = 0.25 in the Sagittarius arm ($V_{\rm LSR}$ from $-$35 km s$^{-1}$ to $-$20 km s$^{-1}$), respectively.  Thus, the ratio is below 0.3 in the other quiescent regions as is consistent with the above estimate for the nearby dark clouds. 

We infer therefore that the present high ratios of 0.7-0.8, in particular, between the $J$=3--2 and $J$=1--0 lines in G347.3--0.5 lend a support for highly excited states rather uncommon among the galactic molecular gas, keeping in mind that a more extensive survey in the CO $J$=3--2 emission over the galactic molecular gas is highly desirable to better establish this in future.  
This temperature rise is perhaps due to local enhanced heating by the SNR, where possible mechanisms may include X-rays, $\gamma$-rays and/or cosmic ray protons.  We shall postpone to deal with more details on the heating mechanisms until the possible contribution by another young source becomes reasonably well understood (see below).

In order to test an alternative possibility to explain the higher temperature we looked for embedded infrared sources towards the CO peaks in the IRAS point sources catalogue (1988).  It turns out that all the three regions are associated with IRAS sources as listed in Table 3 (see also peak A, C, and D in Figure 7) and all of them appear to have steeply-rising far-infrared spectra explicable as embedded young stars.   It is not certain if all these are really at the same distance of the molecular peaks, 1 kpc, and the physical association needs to be more carefully checked.  If we tentatively assume that they are at 1 kpc, their radiation luminosities are estimated as 140-560 $L_{\odot}$, which may contribute to heat up the molecular gas to some degree.  Thus, we shall postpone to affirm the shock heating/compression in the two peaks A and C until better estimates of density and temperature become available and until better establishing of the association of the IRAS sources.

In conclusion, the strong sub-mm emission of CO shown in the present work certainly demonstrate highly excited conditions of the molecular gas interacting with the SNR, while it remains open if the conditions are solely ascribed to the shock interaction or if an additional heating may be caused by young stars embedded.   Further efforts to better constrain the physical parameters of the molecular gas and the nature of the IRAS sources are essentially important.

\subsection{Broad Wings}

Observations made by both of ASTE and CSO indicate that peak C shows fairly strong CO $J$=3--2 emission with some wing-like feature.  The CO $J$=1--0 wings are already discovered and are discussed in terms of the shock acceleration by the blast wave in Paper I.  It is possible that the sub-mm wings indicate the accelerated gas due to the interaction with the SNR. 

Figure 10 shows a CO $J$ = 1--0 profile map of the peak C region.  The broad wings have $\sim$ 20 km s$^{-1}$ extent showing a peak at $-$11 km s$^{-1}$, and are localized within 2-4 arc min ($\sim$ 0.6-1.2 pc at 1 kpc).  The blue-shifted component is seen from $\sim$ $-$23 km s$^{-1}$ to $-$14.5 km s$^{-1}$, and the possible red-shifted component from $\sim$ $-$9.0 km s$^{-1}$ to $-$7 km s$^{-1}$, respectively.  For the red-shifted side, we note that confusion with the other clouds at $V_{\rm LSR}$ from $\sim$ $-$6 km s$^{-1}$ to 0 km s$^{-1}$ makes it difficult to separate the wing component clearly.  There also remains a possibility that the $J$=1--0 red-shifted wing-like profile is not due to the shock acceleration, but just a superposition of unshocked clouds.  
 
As shown in Figure 11, a CO $J$ = 3--2 profile map, peak C exhibits broad wings.   These sub-mm wings are more intense in the blue-shifted side and is more localized than $J$=1--0 wings within $\sim$ 0.3 pc of the intensity maximum of peak C.  In Paper I, it is argued that the CO broad wings in peak C may represent the accelerated molecular gas due to the impacting blast wave of the SNR.   For the blue-shifted wing, we confirm that this is a viable interpretation since the higher density/temperature of the wings indicated by the sub-mm spectrum are consistent with this interpretation. 

The present wings are seen both in the red- and blue-shifted sides of the quiescent gas.  Such a trend is not odd in shocked gas and is seen in the other shock excited cases (e.g., IC 443; van Dishoeck et al. 1993, W28; Arikawa et al. 1999).  These two velocity components may be due to complicated geometry of the shock fronts; a possibility here is that the far and near sides of peak C are being shocked to produce the two components. It is also remarkable that the size of the $J$=3--2 wing is much smaller than that of the $J$=1--0 wings. This suggests that the higher excitation condition in density is localized toward the central part of peak C.  

Here, we should note an alternative possibility that the wings may represent molecular outflow driven by young protostars (for molecular outflows see e.g., Lada 1985; Fukui et al. 1986; 1989; 1993).  The above-mentioned IRAS source IRAS 17089--3951 toward peak C (Table 3) may lend support to this alternative.  The total molecular mass of peak C is estimated to be $\sim$ 400 $M_{\odot}$ by assuming the X-factor for $^{12}$CO($J$=1--0), 2 $\times$ 10$^{20}$ cm$^{-2}$ K$^{-1}$ (km s$^{-1}$)$^{-1}$ (Bertsch et al. 1993), and the luminosity of the IRAS source $\sim$ 300 $L_{\odot}$ seems consistent with a young stellar object embedded in it.   By assuming that it is a protostellar molecular outflow, the physical quantities, kinetic energy, momentum and mass of outflow, are estimated (Table 4) and it seems that these quantities are not unreasonably different from typical parameters of outflow (Fukui et al. 1989; 1993).  

If this interpretation is correct, peak C may represent a pre-existing dense star-forming cloud core which happened to be close to the SN and the enhanced density in it may have allowed the core to survive against the blast wave.  
The possibility of the protostellar outflow in peak C should be further pursued by mid-infrared imaging and spectroscopy of the IRAS source as well as by higher resolution molecular observations.

In summary, we confirm the CO broad wings in peak C both in the $J$=1--0 and $J$=3--2 CO transitions and present two alternative interpretations; one is the shock acceleration by the SNR blast waves as in Paper I and the other is the molecular outflow driven by a protostar.  We need to accumulate more observations to discern these alternatives.

\subsection{Local Cloud Absorption}

The previous comparison of 0$-$10 km s$^{-1}$ velocity window (Figure 2(q)) suggests absorption of soft X-rays may be affecting the apprearance of X-ray distribution.  Generally speaking, the southern region of the SNR is relatively weak in X-ray, suggesting possible absorption; a most notable soft X-ray depression in the ROSAT image is a straight feature extending from ($l$, $b$) $\sim$ (347$\arcdeg$.0, $-$0$\arcdeg$.7) to (347$\arcdeg$.4, $-$0$\arcdeg$.9) over half a degree which may be due to absorption.  In the following, we make a detailed comparison of CO with soft X-ray and show that the soft X-ray image may be considerably affected by the foreground absorption by molecular gas.

Figure 12 is for comparison with the molecular clouds in a velocity range from 6.5 km s$^{-1}$ to 7.5 km s$^{-1}$ that is probably responsible for the absorption of X-rays.  

The main component of the molecular clouds is located in the southern part of the SNR, from $l$ = 346$\arcdeg$.9 to 347$\arcdeg$.4 and from $b$ = $-$0$\arcdeg$.4 to $-$1$\arcdeg$.0.   Including this, we have named seven CO peaks from a through g as listed in Table 5.

We list the likely candidates for absorption below.

1) The main cloud including peaks a, b, c, and e corresponds to the weaker X-ray intensity.  In particular we note that the most intense X-ray arc in the northwest toward the southwest appears bound by the main cloud at ($l$, $b$) $\sim$ (347$\arcdeg$.05, $-$0$\arcdeg$.45).  This suggests that the sharp cut off of the X-ray arc may be due to absorption at least in part.

2) The straight X-ray depression feature from ($l$, $b$) $\sim$ (347$\arcdeg$.0, $-$0$\arcdeg$.5) to (347$\arcdeg$.4, $-$0$\arcdeg$.9) well corresponds to the CO ridge in the same direction.  

3) A few CO peaks are also located towards the regions of weaker X-ray intensity; f at ($l$, $b$) $\sim$ (347$\arcdeg$.55, $-$0$\arcdeg$.2 - +0$\arcdeg$.2) and g at ($l$, $b$) $\sim$ (347$\arcdeg$.55, $-$0$\arcdeg$.75).  The exception is d at ($l$, $b$) = (347$\arcdeg$.20, $-$0$\arcdeg$.25) where a weak CO peak is overlapped with the part of the northwestern X-ray rim. 

These are promising candidates for absorption features and could be more firmly established through detailed spectral analysis in the X-ray energy spectra.  The CO integrated intensities of these peaks are 6.0-12.6 K km s$^{-1}$, corresponding to the atomic column density of $\sim$ 2.8-5.0 $\times$ 10$^{21}$ cm$^{-2}$ (X-factor of 2 $\times 10^{20}$ cm$^{-2}$ K$^{-1}$ (km s$^{-1}$)$^{-1}$ is adopted).  The absorption column density based on the recent X-ray observations (Cassam-Chenai 2004b, Hiraga et al. 2004) ranges 4-10 $\times$ 10$^{21}$ cm$^{-2}$, fairly consistent with our results within a factor of $\sim$ 2. 
This column density of several $\times$ 10$^{21}$ cm$^{-2}$ is large enough to cause the absorption (e.g., Tatematsu et al. 1990), while the probable variation in the X-ray intensity may make it difficult to estimate the amount of absorption quantitatively.  In conclusion, we have shown that the soft X-ray image of the G347.3--0.5, particularly in the south, is likely affected by the foreground absorption due to the local molecular cloud within several 100 pc of the sun, and any detailed analysis of the soft X-ray distribution is to be made by taking this into account.  More quantitative study of the absorption toward the molcular clouds is desirable by using new X-ray datasets in future.  We also need to consider further the possible contribution of H{\small \,I}.  This is to be done in future with appropriate high resolution  H{\small \,I} measurements. 

Anyway, it is likely that most of these clouds in positive velocity are local components within several 100 pc of the sun.  The fact that the clouds can be recognized as dark extinction features in optical pictures (Digitized Sky Survey archive data from ESO/ST-ECF Science Archive) supports this suggestion.

It may be worthy to discuss on the origin of the small positive velocities of the local clouds.  
One of the possible explanations is that these clouds are parts of the Gould belt.  The Gould belt passes through the area of $b$ $\sim$ 0$\arcdeg$-10$\arcdeg$ at $l$ $\sim$ 347$\arcdeg$ (Taylor et al. 1987), close to the location of the clouds in issue.  Taylor et al. (1987) also estimated the distance and the expansion velocity of the Gould's belt ring to be $\sim$ 300 pc and $\sim$ 5 km s$^{-1}$ respectively, which are not inconsistent with the properties of these clouds.  We can speculate other explanations for the positive velocities, such as the random motion or the streaming motion of the clouds.  Dispersion due to random motion of local clouds is measured in order of seveal km s$^{-1}$ (e.g., $\sim$ 4 km s $^{-1}$ by Liszt et al. 1984,  $\sim$ 8 km s $^{-1}$ by Stark \& Brand 1989), and the streaming motion due to the velocity field of the Galactic density wave is estimated to be $\sim$ 4 km s$^{-1}$ in the solar vicinity (e.g. Burton \& Bania 1974).  Each of these hypotheses is capable explaining of the positive velocity of $\sim$ 6--7 km s$^{-1}$ in this case.


\section{Discussion}

\subsection{Distance}

Based on the good spatial correlation between CO and X-ray the authors in Paper I argued that the distance of the SNR G347.3--0.5 is most likely 1 kpc instead of 6 kpc previously favoured.  The present detailed analysis of the same CO datasets also confirms this small distance. 

The peak velocity range of the molecular clouds named A-Y in section 3.3. is $-$12 - $-$3  km s$^{-1}$ (Table 1).  Most of the clouds are highly likely to be in almost the same distance and associated with the SNR, since they show good spatial coincidences with the X-ray distribution. 
The velocity centroid of these clouds is $\sim$ $-$6 km/s as remarked in Paper I, giving the kinematic distance of $\sim$ 1 kpc.  Here we adopt the Galactic rotation curve model by Brand \& Blitz (1993) for deriving kinematic distances. 
If one assumes that the distance of these clouds in the velocity range of $-$12 - $-$3  km s$^{-1}$ are uniformly 1 kpc, they are localized in an area of $\sim$ 20 pc (= 1$\arcdeg$ in angular size) and have a dispersion of the peak velocities of $\sim$ 9 km/s in maximum each other.
A simple and feasible explanation of this cloud-cloud velocity dispersion is the initial proper motion.  Generally, it is not unusual that localized molecular clouds have velocity dispersions of $\sim$ several - 10 km s$^{-1}$, being due to random and/or streaming motion.  Thus, in this case it is also natural to regard these clouds in the velocity range of $-$12 - $-$3 km s$^{-1}$ as having almost the same distance of $\sim$ 1 kpc.

We note that the range of the kinematic distance corresponding the velocity range of -12 - -3 km s$^{-1}$  is about 0.5 - 2.0 kpc.  However, such an uncertainty does not conflict with our supposition that the G347.3--0.5 is at the small distance basically.

It is also noteworthy that recent work, independent of Paper I, have provided support for the small distance; these are studies of the X-ray and interstellar absorption (Uchiyama et al. 2003; Koo et al. 2003; Cassam-Chenai et al. 2004b), H.E.S.S. TeV $\gamma$-ray image (Aharonian et al. 2004).  Koo et al. (2003) and Cassam-Chenai et al. (2004b) made analyses of the interstellar absorption by using low-resolutional H{\small \,I} (15$'$--30$'$) and CO (8$'$) data and concluded that the X-ray absorption is consistent with the small distance around 1 kpc as opposed to the large distance around 6 kpc. Uchiyama et al. (2003) posed a constraint that the shock velocity must be $>$ 5000 km s$^{-1}$ from Chandra observations of the northwestern shell, implying that G347.3-0.5 is a quite young SNR.  The physical properties of the SNR derived assuming the distance of 1 kpc is consistent with this demand for the shock velocity (see next subsection). 
The H.E.S.S. TeV $\gamma$-ray image has revealed a detailed distribution of the highest energy photons.  Their major finding is the shell-like distribution of TeV $\gamma$-ray similar to the X-ray distribution.   These authors support the distance 1 kpc by noting that the TeV $\gamma$-ray peaks are stronger in the northwestern region where the interaction between the local molecular gas and the SNR is suggested in Paper I and that the TeV $\gamma$-ray flux is weak toward the eastern regions where cloud A was suggested to be interacting.

The small distance 1 kpc is therefore acquiring even stronger supports since the publication of Paper I.

\subsection{Evolution of the SNR}

In this section we discuss the evolution of the SNR specifically for the case of 1 kpc distance which is strongly supported observationally by the present work as well as by others (e.g., Koo et al. 2003, Cassam-Chenai et al. 2004b, Aharonian et al. 2004).  Slane et al. (1999) already discussed that the two distances, 1 kpc and 6 kpc, are both possible in explaining the evolution of the SNR though the time scales etc. should be quite different between the two cases.  We shall here highlight the main aspects of the SNR evolution at 1 kpc; we do not intend to use the evolutionary models to discern the distance since it is no more required. 

There are seven historical supernovae and G347.3--0.5 likely corresponds to the historical SN AD 393, the only one that has escaped identification among the seven recorded in the historical literature (Wang et al. 1997).  Thus, we confirm that G347.3--0.5 corresponds to SN AD 393 and adopt its age as 1600 yrs.  

The small distance 1 kpc drastically changes many physical parameters of the SNR from those expected in case of 6 kpc.   Various physical parameters are compared for the two distances 1 kpc  and 6 kpc in Table 6.  At 1 kpc, the radius of the SNR is 8.7 pc and the age is as small as $\sim$ 1600 yrs.  Because of lack of thermal X-ray emission, the emission measure of the X-ray emitting hot gas should be very small, like $\sim$ 10$^{15}$ cm$^{-5}$ (e.g., Slane et al. 1999).  Such a very low emission measure should require very low densities of the post-shock region.   It also indicates that the total mass of swept-up matter is not large enough to achieve the adiabatic phase and the remnant is still in free-expansion phase with thermal non-equilibrium between electrons and ions.  If we assume that the progenitor is a massive star and the ejecta mass is 3 $M_{\odot}$ (Borkowski et al. 1996), the explosion energy of 10$^{51}$ ergs gives the ejecta velocity of 5,800 km.   This velocity is nearly equal to average velocity of the shock front $\sim$ 5,500 km s$^{-1}$ as derived from X-ray measurements (e.g., Koyama et al 1997, Uchiyama et al. 2003), indicating that the blast-wave has not been decelerated yet.  
If we apply a simple model that the free expansion phase ends when the swept-up mass becomes equal to the mass of ejecta, we infer that the explosion occurred in low-density surroundings having density of $<$ 0.01 cm$^{-3}$ and that the ejecta is non-radiative at present.  We then obtain emission measure of $\int n_{\rm e} n_{\rm H}$ $dl$ $<$ 3 $\times$ 10$^{15}$ cm$^{-5}$.  This value is nearly the same as that in case of 6 kpc distance (Table 6) and is small enough to explain the lack of thermal emission.   In addition, the electron temperature is estimated to be $<$ 0.5 keV by applying the analytic formula developed by Masai (1994) to the X-ray data (Slane et al. 1999).  This low value of electron temperature is consistent with the observed X-ray spectrum.  We confirm that the basic X-ray properties of the SNR are explained consistently at 1 kpc.

Slane et al. (1999) reported that the spectrum of the central object 1WGA J1713.4--3949 can be fitted with 0.38 keV blackbody model and the radius of emitting region is 0.5$^{+0.16}_{-0.11}$ $\times$ diameter at 1 kpc.  The temperature of the blackbody model is significantly higher than that of a cooling neutron star whose age is $\sim$ 1600 year (e.g., Slane et al. 2002).  The polar cap heating model seems to be good in this case, and is able to explain the observed temperature and the radius of the emitting region plausibly at 1 kpc.  Here, we have to notice another possibility still remaining that the central object is not associated with the SNR.  Slane et al. (1999) for instance pointed out that the X-ray spectra of the central object can be fitted either by a  power-law or thin thermal emission models with absorption having the column density comparable to the total absorption through the Galaxy in this direction. 
To summarize, the physical parameters are well explained in the framework of very young SNR in the free expansion phase at 1 kpc.  Another detailed discussion on the SNR evolution may be found elsewhere (e.g., Cassam-Chenai et al. 2004b).

\subsection{Origin of Gamma-Rays and Cosmic Ray Protons}

Higher quality multi-wavelength data including radio, X-ray and $\gamma$-ray are needed to obtain definitive conclusions on cosmic ray acceleration in G347.3--0.5, the most unique object known to date in testing the proton acceleration.  Higher energy X-ray imaging observations will provide synchrotron cut-off energy and more sensitive X-ray spectroscopy is needed to detect thermal X-ray emission.  $\gamma$-ray observations with higher spatial resolution will allow us to better constrain the acceleration mechanism.

It is in this context a very important issue to understand precisely the mechanism of TeV $\gamma$-ray emission in G347.3--0.5.  For this purpose, it is essential to discern the contributions of the cosmic ray proton component from the electronic contribution via inverse Compton process and this should be achieved by making a detailed comparison between spatially resolved images of CO and TeV $\gamma$-ray.  

The recent H.E.S.S. image of TeV $\gamma$-ray (Aharonian et al. 2004) is certainly encouraging in that it is beginning to resolve the TeV-$\gamma$ distribution at a few arc min, similar to the angular resolution of NANTEN CO dataset.  It is particularly intriguing that the TeV $\gamma$-ray distribution largely resembles that of the CO distribution in that they are enhanced in the northwestern rim of the SNR as noted by Aharonian et al. (2004).   While the statistical significance in H.E.S.S. results may not yet be high enough to make a detailed comparison of the individual peaks, we make a preliminary estimate of the efficiency for proton acceleration over the entire SNR by using a formula from Enomoto et al. (2002), ($E$/10$^{48}$)($M_{\rm cloud}$/200)($l$/3)$^{-3}$($d/1)^{-5}$ = 1.35 , where $E$ (erg) is the total energy of cosmic ray protons, $M_{\rm cloud}$ ($M_{\odot}$) the molecular cloud mass interacting with them, $l$ (pc) the typical length of the cloud, and $d$ (kpc) the distance, 
same as that in Paper I.  We assume that the interacting molecular cloud mass amounts to $\sim$ 2500 $M_{\odot}$ by summing up the major CO peaks in contact with the SNR (from Table 1).  The equation gives the total energy of cosmic ray protons of $\sim$ 2.4 $\times$ 10$^{49}$ erg, and then, the efficiency of the proton acceleration is estimated to be $\sim$ 0.024 by dividing by the total energy of the SN explosion, $\sim$10$^{51}$ erg.


\section{Conclusions}

We conclude the main results of the present work as follows;

\begin{enumerate}

\item The present detailed analysis of the NANTEN CO data has established that the SNR G347.3--0.5 is interacting with the molecular gas at a distance of 1 kpc instead of 6 kpc previously suggested.  This is supported by subsequent H{\small \,I} and TeV $\gamma$-ray studies.  The physical picture of the SNR evolution is shown to be consistent with the distance 1 kpc if the SNR is in the free expansion phase.  
\item The intense CO $J$=3--2 emission has been detected toward two of the CO peaks, strongly indicating that the interacting molecular gas is in a highly excited states in density and temperature unusual in typical local dark clouds.  This provides another support for the interaction.  
\item We note an alternative possibility that IRAS point sources, candidates for embedded young stars in the CO peaks, may provide additional heating in these CO peaks, although their actual association needs to be tested by further observations in the infrared and others.  
\item Peak C showing strong interaction with the SNR blast wave is established to exhibit broad CO wings in the $J$=1--0 and $J$=3--2 transitions.  Two alternative interpretations for the wings are presented; one is the shock acceleration (Paper I) and the other protostellar outflow.  Further observational efforts are desirable to discern them.
\item It is demonstrated that the southern part of the SNR may suffer from significant absorption by a foreground molecular gas within several 100 pc, while quantitative estimate for the absorption is not available at present.  

\end{enumerate}


\acknowledgments

We would like to thank all the stuff members of the NANTEN project.  We also acknowledge Prof. Tadayuki Takahashi, Dr. Yasunobu Uchiyama, Dr. Hiraga Junko, for their helpful comments.  We appreciate the hospitality of all the people of Las Campanas Observatory of the Carnegie Institution of Washington, and thank Prof. Tom Phillips for kindly allocating the observing time of the CSO (The Caltech Submillimeter Observatory).  The authors greatly acknowledge the hospitality of all members of NAOJ (National Astronomical Observatory Japan), University of Tokyo, and Osaka Prefecture University who are working for the ASTE project, for thier dedicated supports.  The NANTEN project was based on the mutual agreements between Nagoya University and Carnegie Institution of Washington.  We also acknowledge that NANTEN project was realized by the contributions from many Japanese public donors and companies.

This reserch was financially supported by JSPS Grant-in-Aid for Scientific Research (B) No. 14403001 and MEXT Grant-in-Aid for Scientific Research on Priority Areas No. 15071202.



\begin{figure*}
\epsscale{0.8}
\plotone{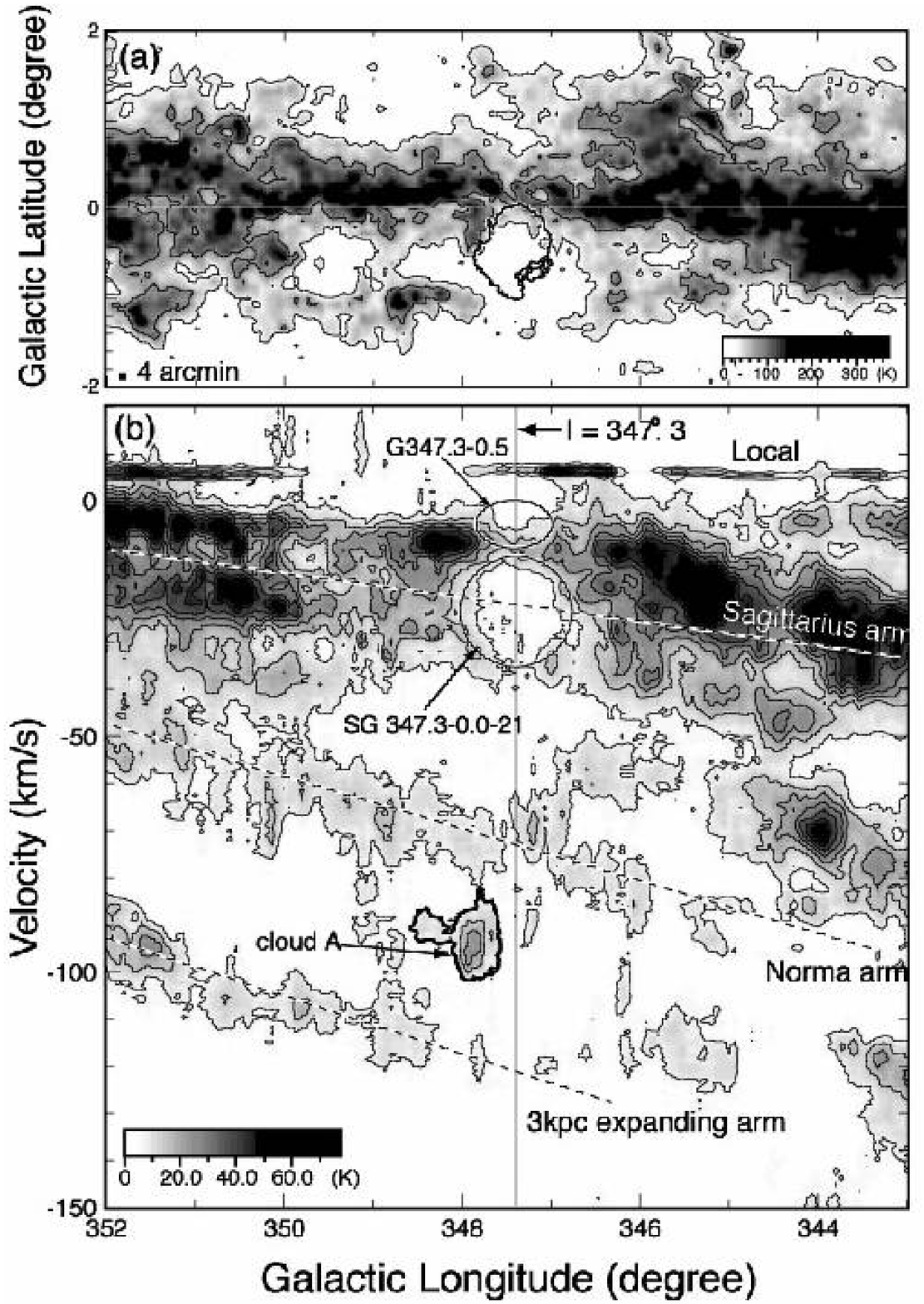}
\caption{(a) A large area distribution of $^{12}$CO($J$=1--0) integrated ntensity taken from NANTEN galactic plane CO survey (Fukui et al 2005; see also Matsunaga et al. 2001).   The grid spacing is 4$\arcmin$ with a 2$\arcmin$.6 beam.   The lowest contour level and the contour interval are 30 K km s$^{-1}$ and 80 K km s$^{-1}$ in $T_{\rm R}^{*}$, respectively.  The X-ray boundary of G347.3--0.5 (Pfeffermann \& Aschenbach 1996) is indicated by thick line (colored red in electronic edition).  (b) A Galactic longitude-velocity map of $^{12}$CO($J$=1--0) in the same region as (a).  The integrated range in Galactic latitude is from $-$1$\arcdeg$ to 1$\arcdeg$.  Both the lowest contour level and interval are 7 K, respectively.  Approximate positions of Sagittarius arm, Norma arm, and 3 kpc expanding arm are indicated by dashed lines.  The position of $l$ = 347$\arcdeg$.3 is denoted by a solid line.  The SNR G347.3--0.5 and the supershell SG 347.3--0.0--21 (Matsunaga et al. 2001) are indicated by a solid line ellipse (colored red in electronic edition) and a circle, respectively.  The boundary of cloud A (Slane et al. 1999) is denoted by a thick solid line.\label{fig1}}
\end{figure*}

\begin{figure*}
\epsscale{0.8}
\plotone{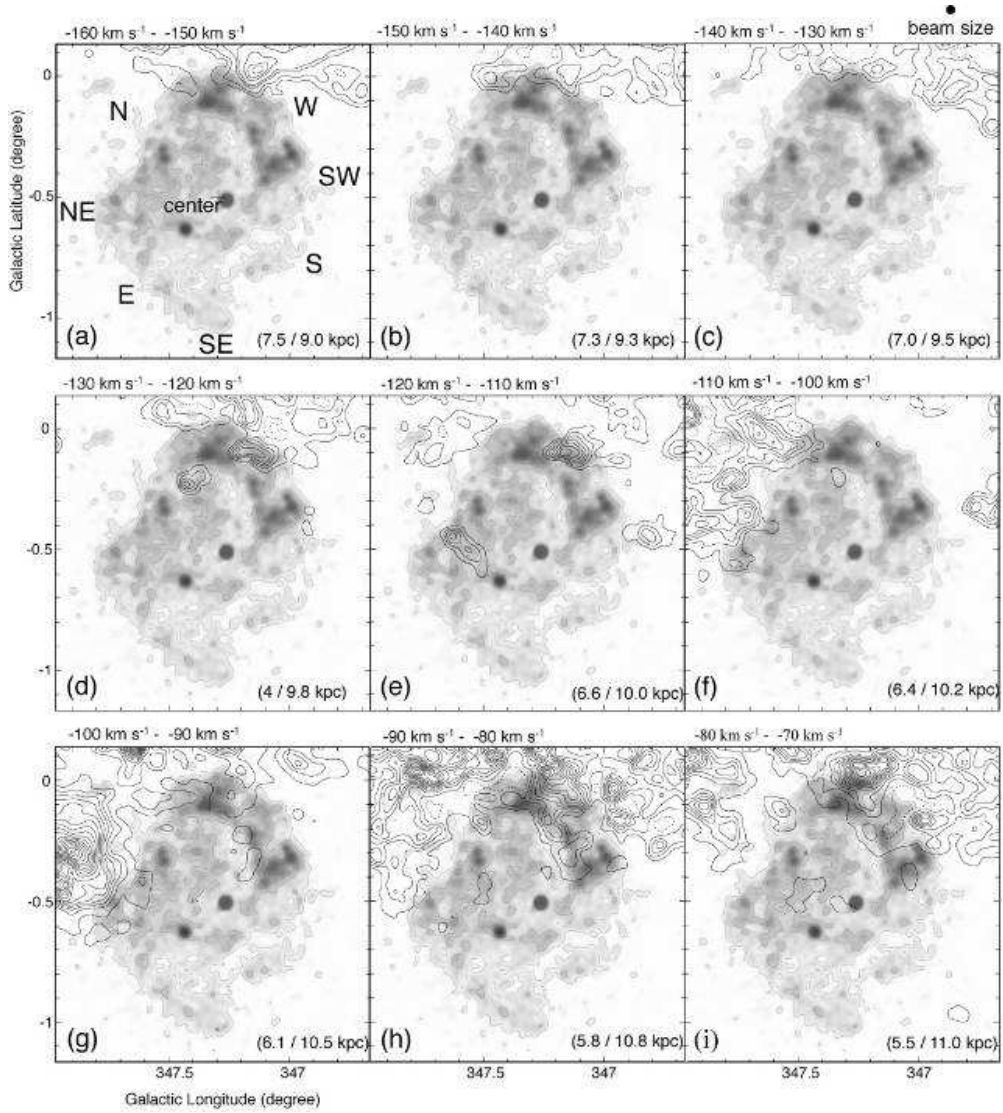}
\setcounter{figure}{1}
\caption{$^{12}$CO($J$=1--0) channel maps (black contours) are overlapped with the X-ray image (gray scale and contours, colored light red in electronic edition) by ROSAT PSPC X-ray Survey (Slane et al. 1999; from ROSAT archive database).  Each panel shows a CO intensity map integrated over the velocity range from $-$160 km s$^{-1}$ to 20 km s$^{-1}$ every 10 km s$^{-1}$,  The minimum contour level and the contour interval are 4 K km s$^{-1}$ in all of the panels, except (g), (i), (j), and (o).  The minimum contour level is 4 K km s$^{-1}$ and the contour interval is 8 K km s$^{-1}$ in (g), (i), (j).  In (o), the contour levels are 4, 8, 12, 16, 20, 24, 32, 40, 48, 56, 64, 72, 80 K km s$^{-1}$, respectively.  The kinematic distances corresponding to $V_{\rm LSR}$ are also indicated in the bottom of each panel.\label{fig2}}
\end{figure*}

\begin{figure*}
\epsscale{0.8}
\plotone{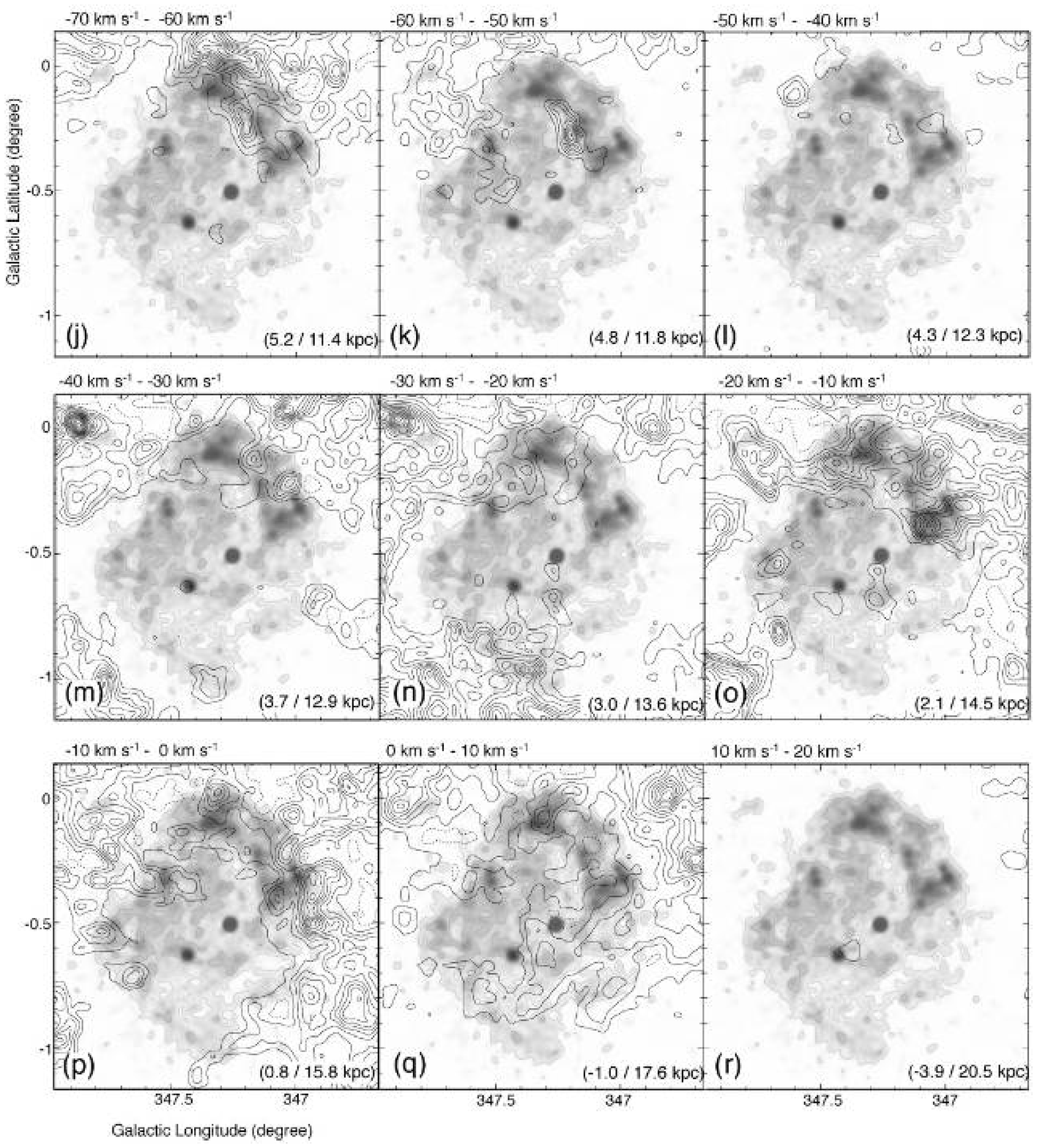}
\setcounter{figure}{1}
\caption{Continued. \label{fig2}}
\end{figure*}

\begin{figure*}
\epsscale{0.8}
\plotone{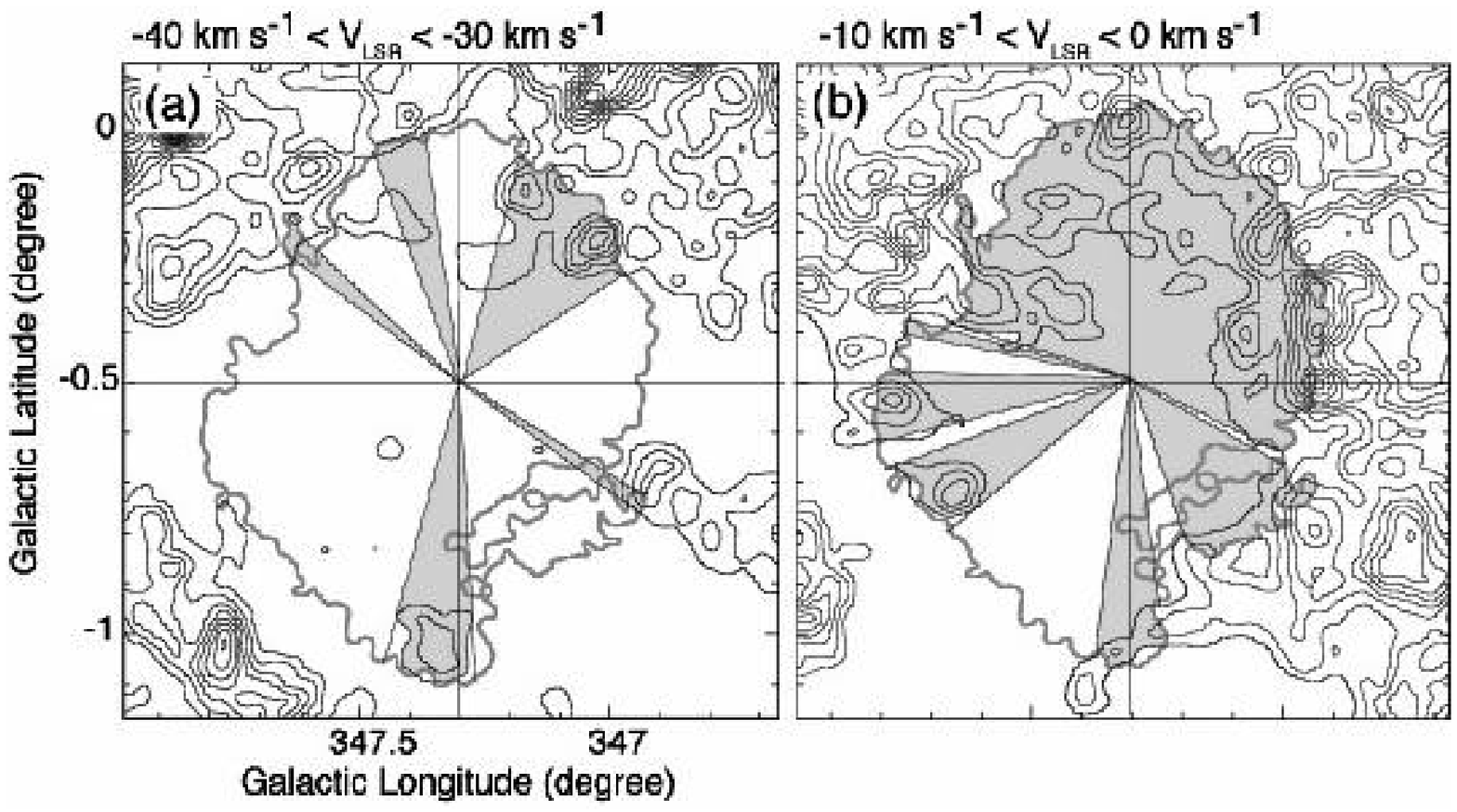}
\caption{A pair of the $^{12}$CO($J$=1--0) velocity channel maps (black contours) overlapped with the ROSAT X-ray boundary line of the SNR (gray contours).  The velocity ranges are from $-$40 km s$^{-1}$ to $-$30 km s$^{-1}$ for (a), and from $-$10 km s$^{-1}$ to 0 km s$^{-1}$ for (b), respectively.  The minimum contour level and the contour interval of CO are 4 K km s$^{-1}$, respectively.  The covering angles (see text) measured from ($l$, $b$) = (347$\arcdeg$.3, $-$0$\arcdeg$.5) are indicated by gray colored regions.  
 \label{fig3}}
\end{figure*}

\begin{figure*}
\epsscale{0.5}
\plotone{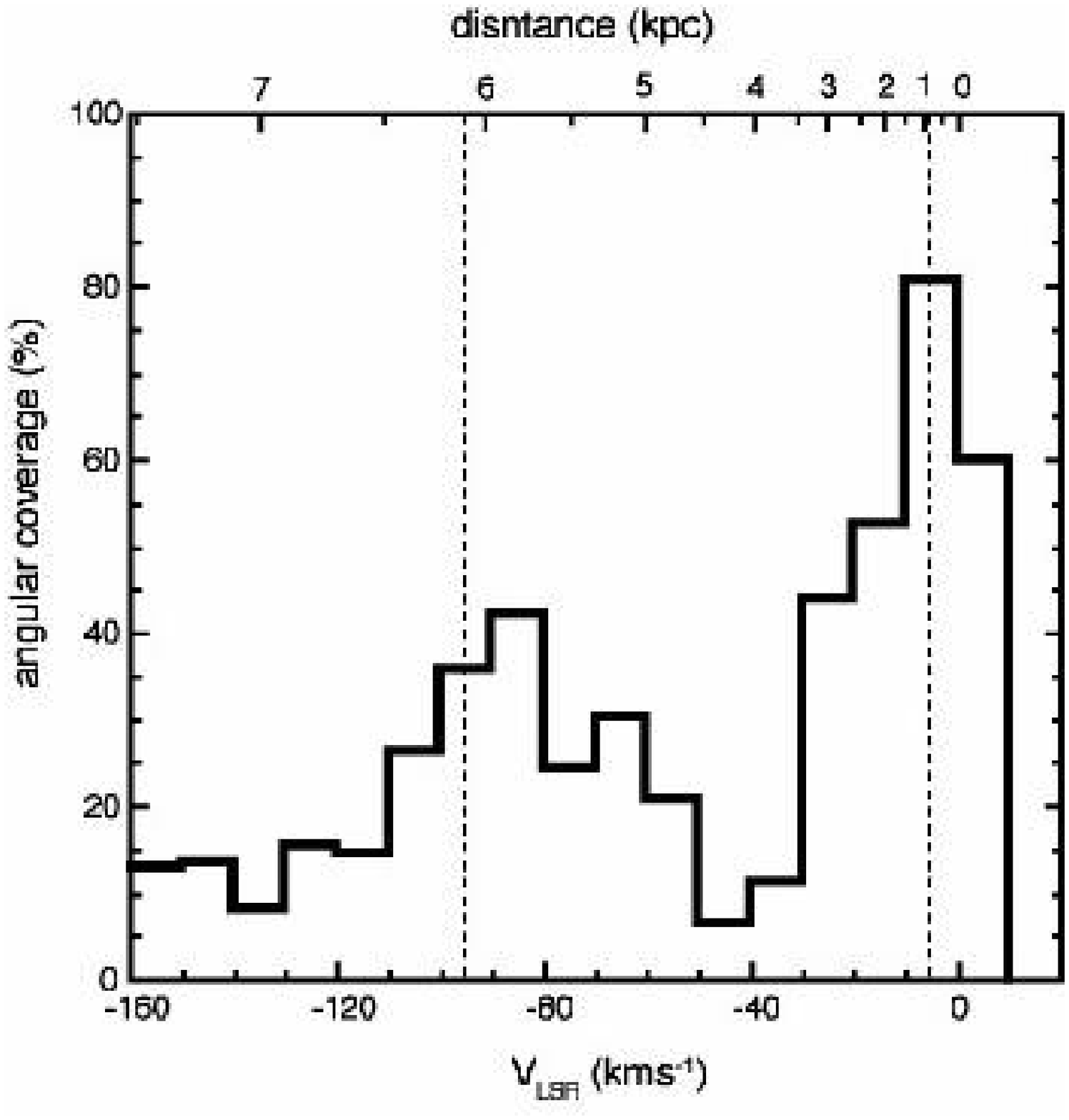}
\caption{A histogram of the angular coverage of CO distribution surrounding the boundary of the SNR (see Appendix A for details).   The covering factor is defined as a fraction of the angle summed up, where the CO distribution is overlapped with the outer boundary of the X-ray SNR.  The mean central position of the X-ray emitting area is taken as ($l$, $b$) = (347$\arcdeg$.3, $-$0$\arcdeg$.5), and for each velocity bin of 10 km s$^{-1}$, the CO boundary is defined as the lowest contour level in each channel map of Figure 2.  The kinematic distance scale corresponding to $V_{\rm LSR}$ is also indicated in the top of each panel, and the assumed ditances of G347.3--0.5 (1 kpc/6 kpc) are denoted by dashed lines.\label{fig4}}
\end{figure*}

\begin{figure*}
\epsscale{1}
\plotone{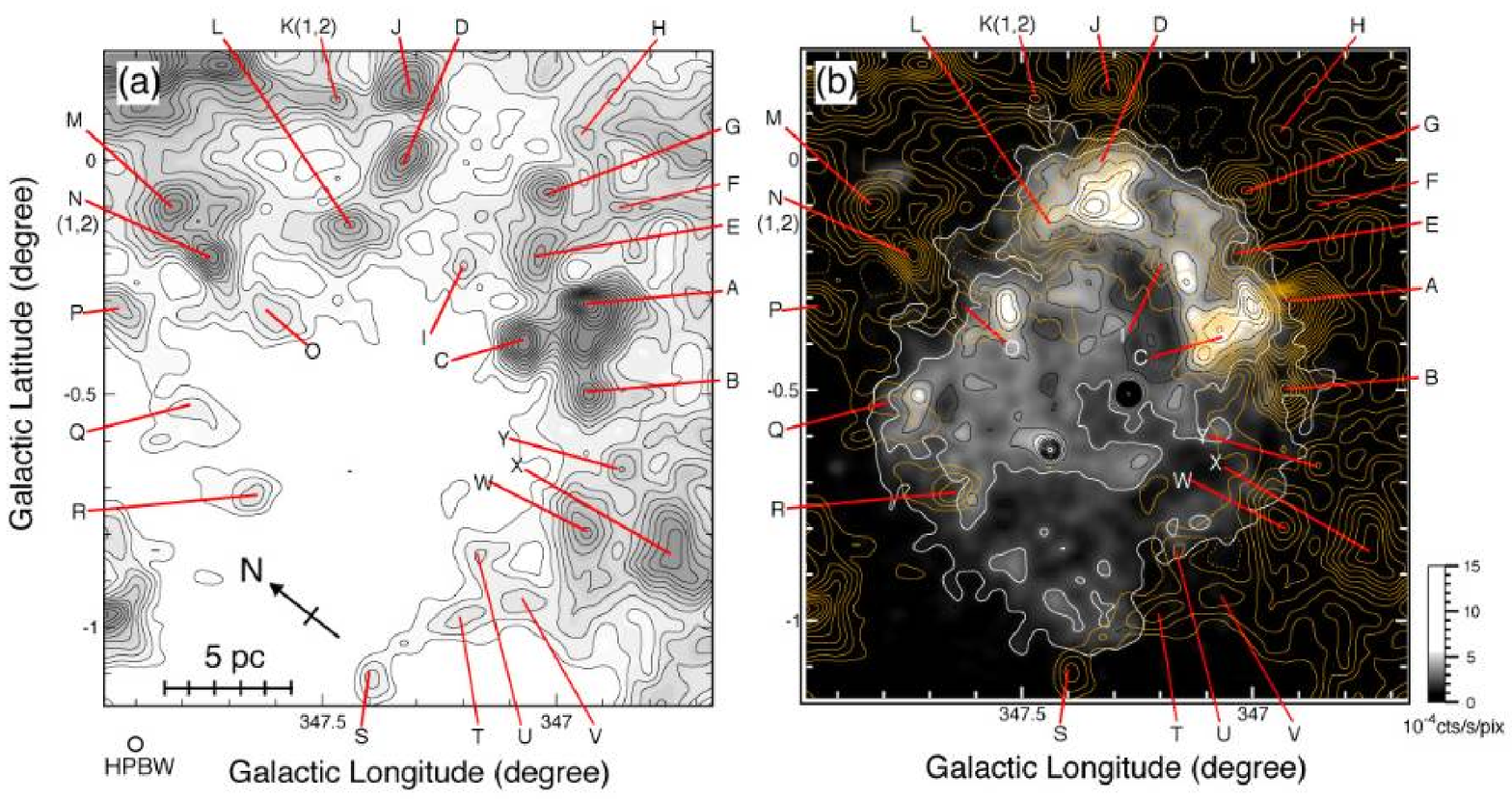}
\caption{(a) An intensity distribution of $^{12}$CO($J$=1--0) emission.  The intensity is derived by integrating the $^{12}$CO($J$=1--0) spectra from $-$12 km s$^{-1}$ to $-$3 km s$^{-1}$.  The lowest contour level and interval are 2.5 K km s$^{-1}$.  The CO peaks A$-$Y discussed in the section 3.3 are indicated in the figure.  (b) An overlay map in Galactic coordinates showing G347.3--0.5 X-ray image in gray scale (from ROSAT archive database) and the $^{12}$CO($J$=1--0) distribution in white contours (colored yellow in electronic edition).  The contour levels and the velocity range of CO map are the same with those of Figure (a).
\label{fig5}}
\end{figure*}

\begin{figure*}
\epsscale{1}
\plotone{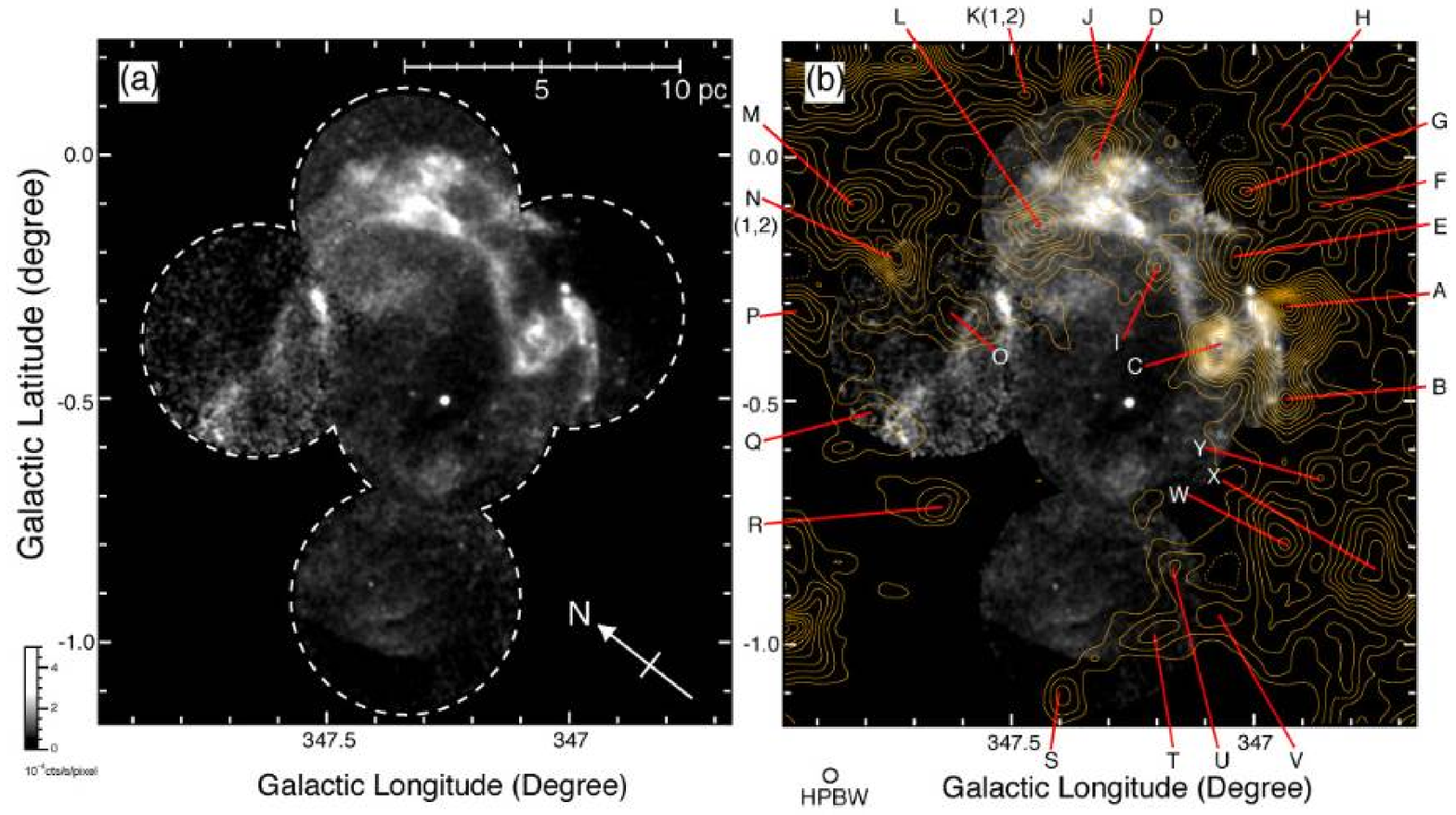}
\caption{(a) An X-ray image of G347.3--0.5 taken by XMM (Hiraga et al. 2004) in gray scale.  An observed area is indicated with dashed lines.  (b)An overlay map of G347.3--0.5 XMM image in Figure (a) and $^{12}$CO($J$=1--0) intensity contours (colored yellow in electronic edition).  The CO contours and the velocity range are the same with those in Figure 5.  The depressions of the CO emission with the lowest contour level inside of the molecular boundary are shown with dashed contours to avoid confusion.  The CO peaks A$-$Y discussed in the section 3.3 are indicated. \label{fig6}}
\end{figure*}

\begin{figure*}
\epsscale{0.9}
\plotone{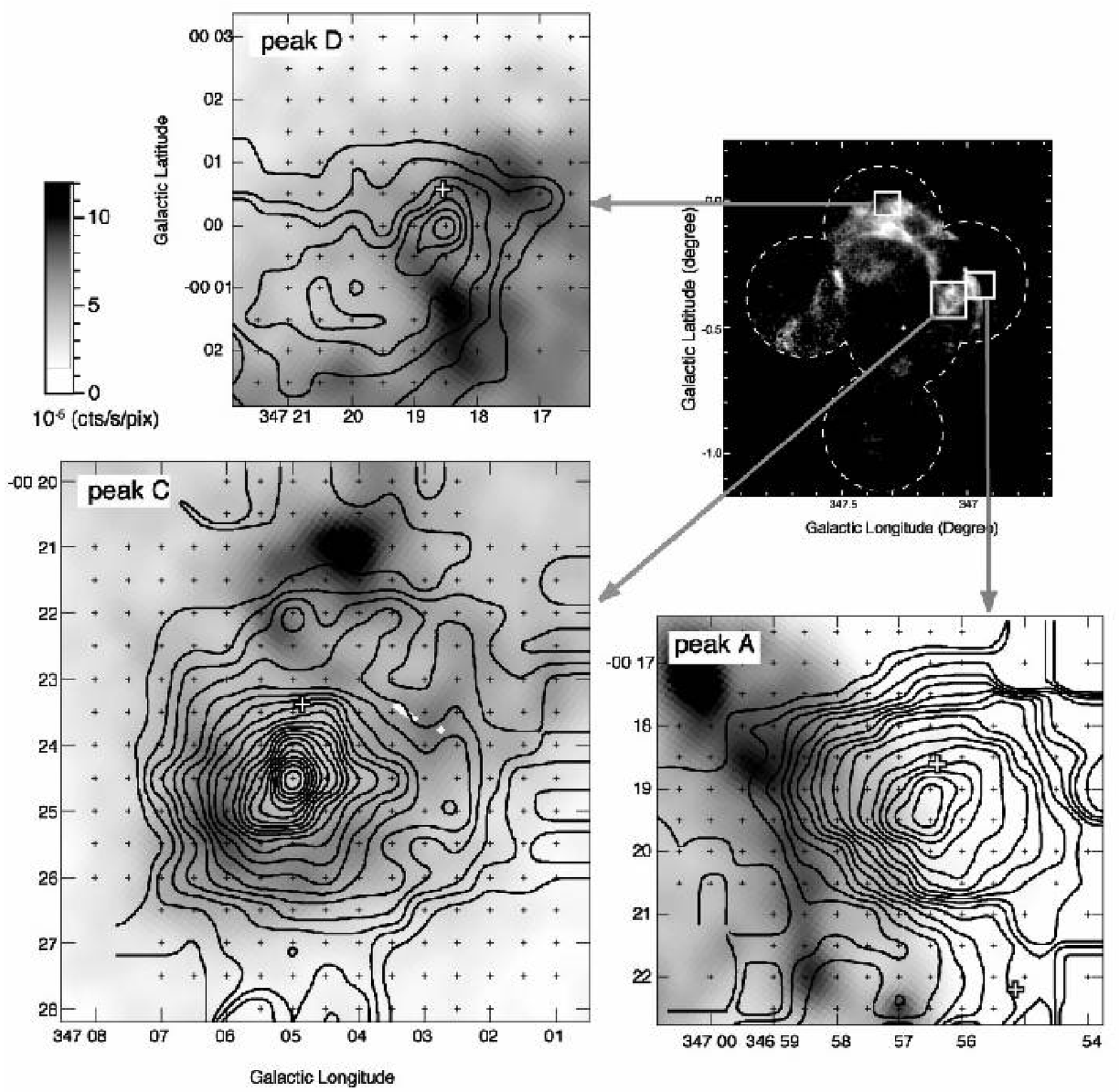}
\caption{The distribution of the CO ($J$=3--2) emission obtained with the ASTE sub-mm telescope (contours) are overlapped with the XMM X-ray image (gray scale).  The beam size of the ASTE was 23$\arcsec$.  The three regions including CO peaks A, C and D are shown with an insert of the XMM image on the top right.  The velocity range is from $-$12 km s$^{-1}$ to $-$3 km s$^{-1}$ in peak A and peak C, from $-$12 km s$^{-1}$ to $-$8 km s$^{-1}$ in peak D.  The CO contours are every 1.5 K km s$^{-1}$ from 3.0 K km s$^{-1}$ in peak C, and peak D, every 1.5 K km s$^{-1}$ from 6.0 K km s$^{-1}$ in peak A respectively.  Observed positions are indicated by crosses.  Open crosses show the positions of the IRAS point sources in each panel (see Table 3).  The white boxes in the X-ray image at ($l$, $b$) = (347$^{\circ}$03$^{\prime}$, $-$00$^{\circ}$23$^{\prime}$.5) are due to the absence of the data. 
\label{fig7}}
\end{figure*}

\begin{figure*}
\epsscale{0.5}
\plotone{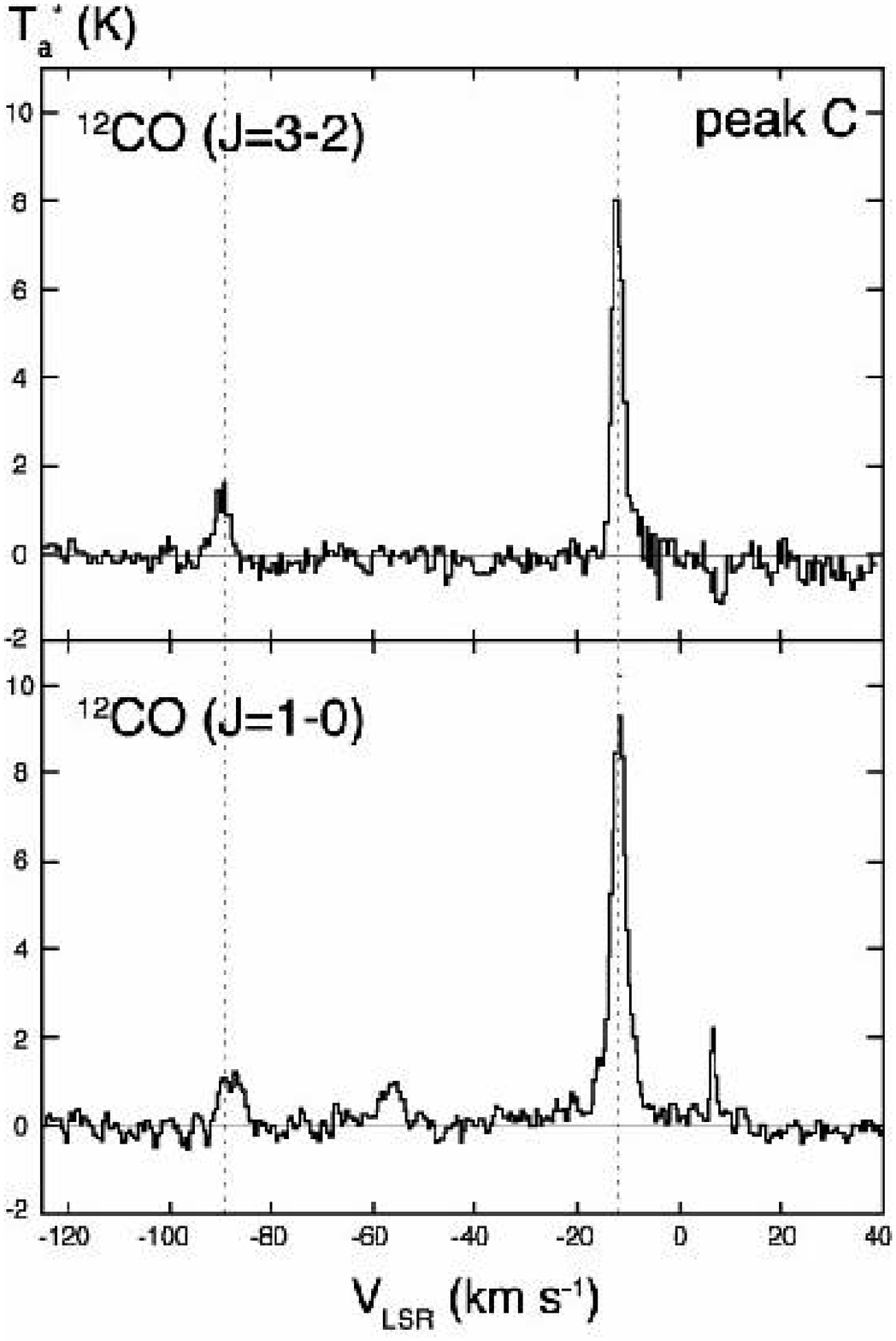}
\caption{Upper: The $^{12}$CO($J$=3--2) spectrum at ($l$, $b$) = (347$\arcdeg$.100, $-$0$\arcdeg$.400) covering both of $-$11 km s$^{-1}$ component and $-$90 km s$^{-1}$ one, taken by CSO.  Lower: The $^{12}$CO($J$=1--0) spectrum taken with NANTEN at ($l$, $b$) = (347$\arcdeg$.100, $-$0$\arcdeg$.367).   The peak velocities of $-$11 km s$^{-1}$ component and $-$90 km s$^{-1}$ component are denoted by dashed lines. \label{fig8}}
\end{figure*}

\begin{figure*}
\epsscale{0.5}
\plotone{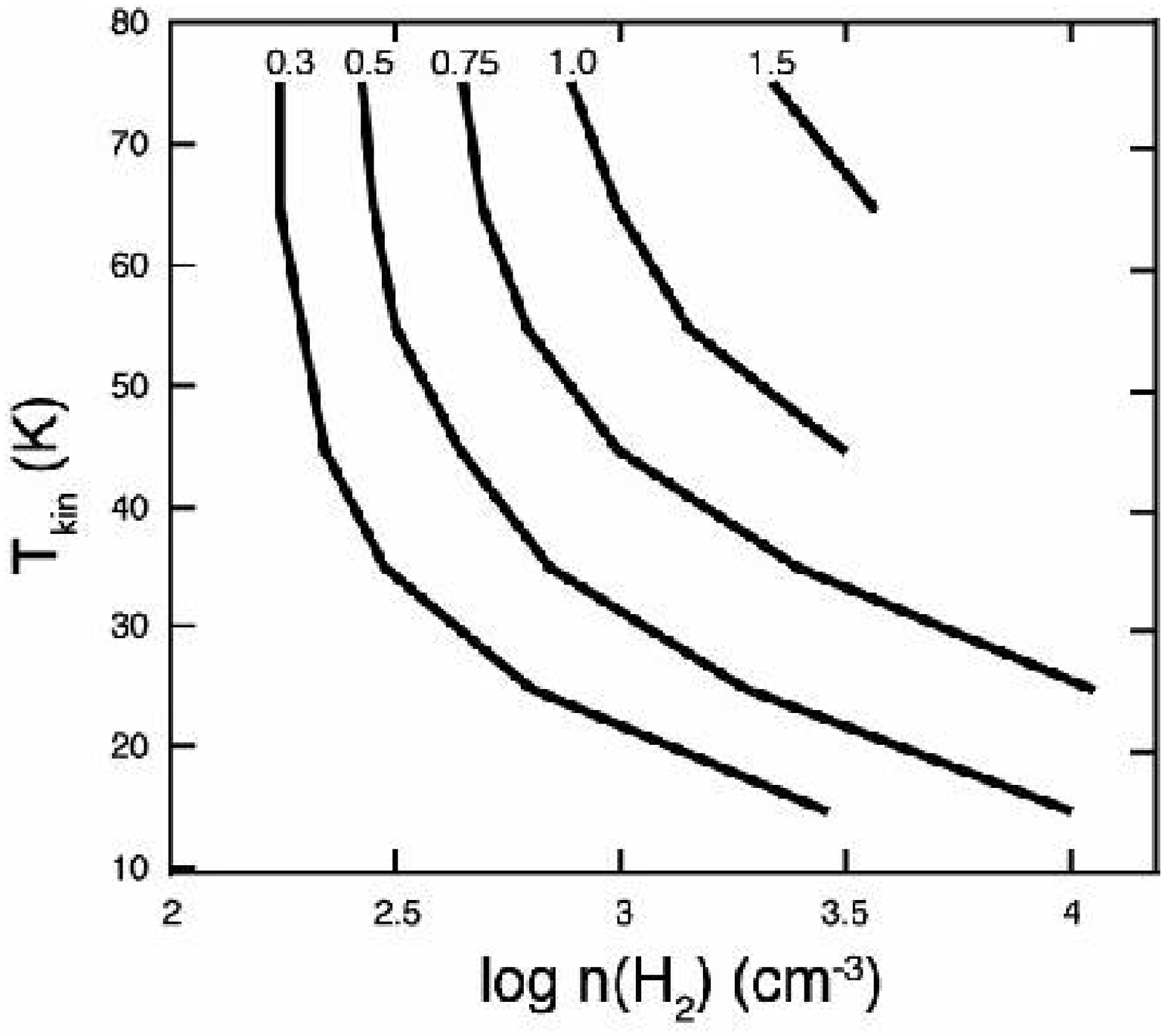}
\caption{The results of LVG caluculations for $X$(CO)/$V$ = 10$^{-4.5}$ pc km$^{-1}$ s.  The calculations are carried out for the value of $T_{\rm kin}$ from 15 K to 75 K with an interval of 10 K.  Solid lines indicate constant values of $T_{\rm R}^{*}$($J$ = 3--2)/$T_{\rm R}^{*}$($J$ = 1--0) = 0.3, 0.5, 0.75, 1.0, and 1.5, respectively. 
\label{fig9}}
\end{figure*}

\begin{figure*}
\epsscale{0.7}
\plotone{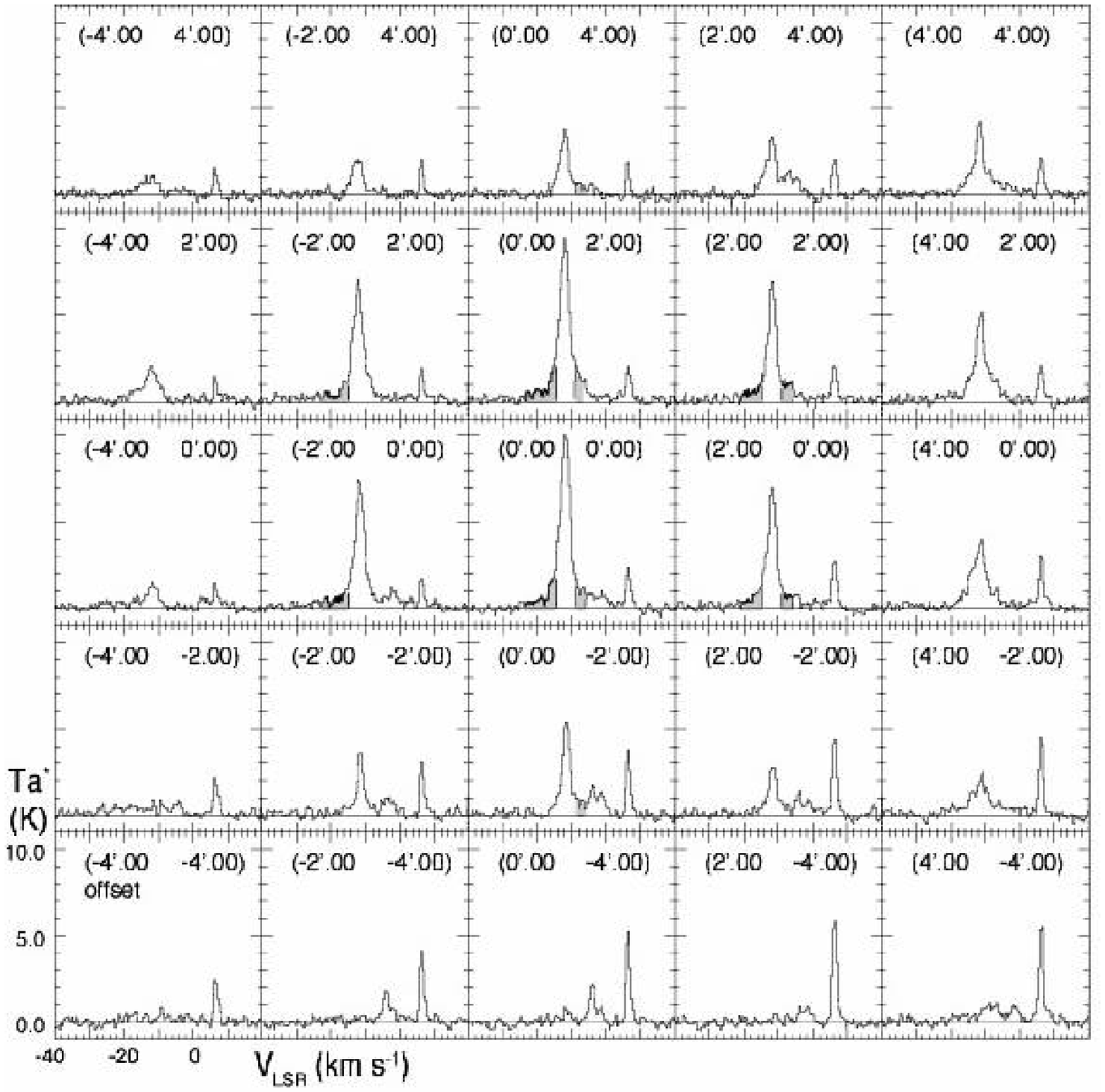}
\caption{A profile map of $^{12}$CO($J$=1--0) spectra toward peak C.   The central position is ($l$, $b$) = (347$\arcdeg$.07, $-$0$\arcdeg$.40) and the offset for each position is 2$\arcmin$ in the Galactic coordinates.  The offset values in arcminutes are denoted in each panel.  The velocity range is from $-$40 km s$^{-1}$ to 20 km s$^{-1}$.  The wing components are indicated by gray colored areas. 
\label{fig10}}
\end{figure*}

\begin{figure*}
\epsscale{0.7}
\plotone{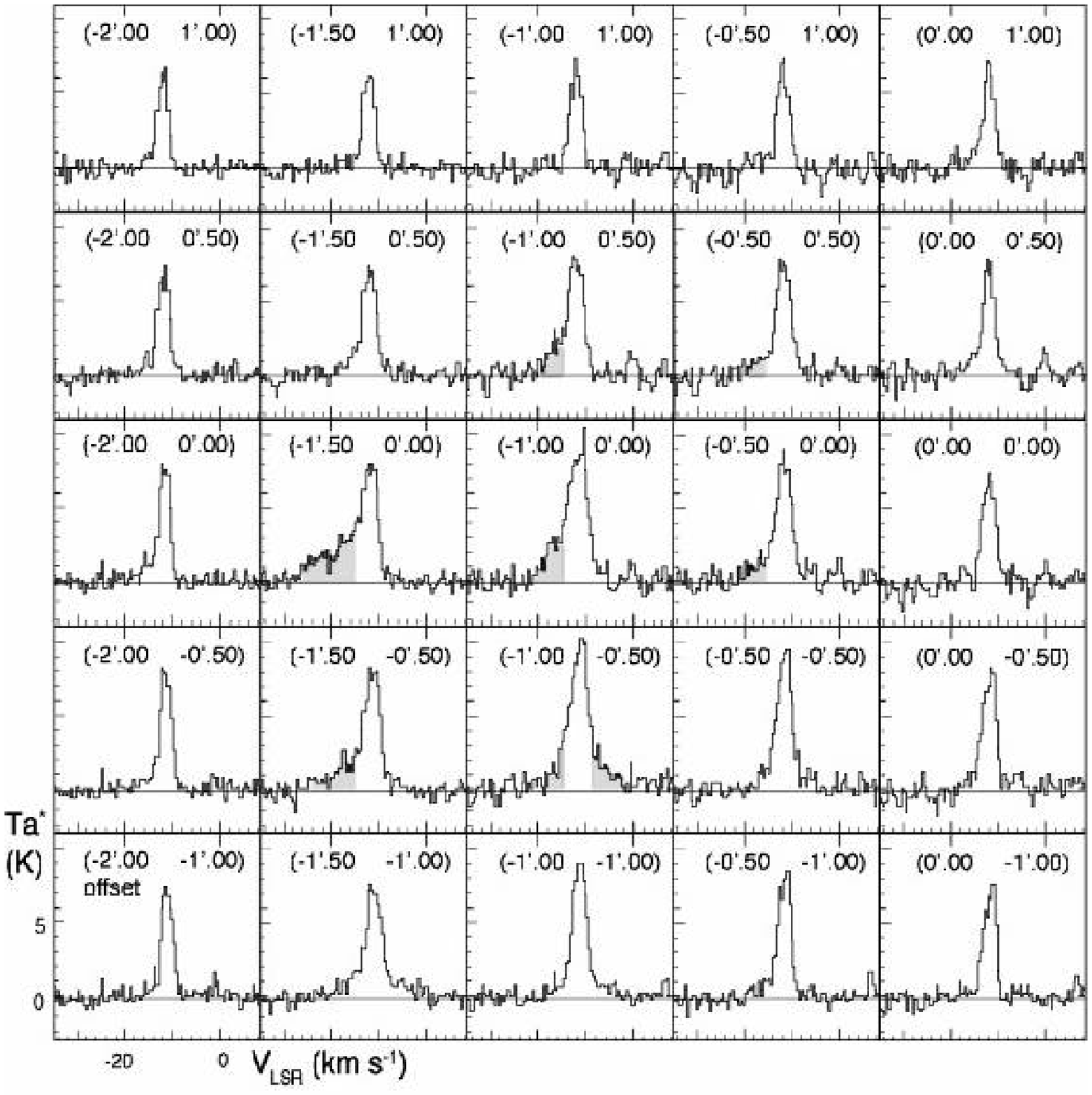}
\caption{A profile map of $^{12}$CO($J$=3--2) spectra toward peak C.   The central position is ($l$, $b$) = (347$\arcdeg$.08, $-$0$\arcdeg$.400) and the offset for each position is 0$\arcmin$.5 in the Galactic coordinates.  The offset values in arcminutes are denoted in each panel.   The velocity range is from $-$35 km s$^{-1}$ to 8 km s$^{-1}$.  The wing components are indicated by gray colored areas. \label{fig11}}
\end{figure*}

\begin{figure*}
\epsscale{0.6}
\plotone{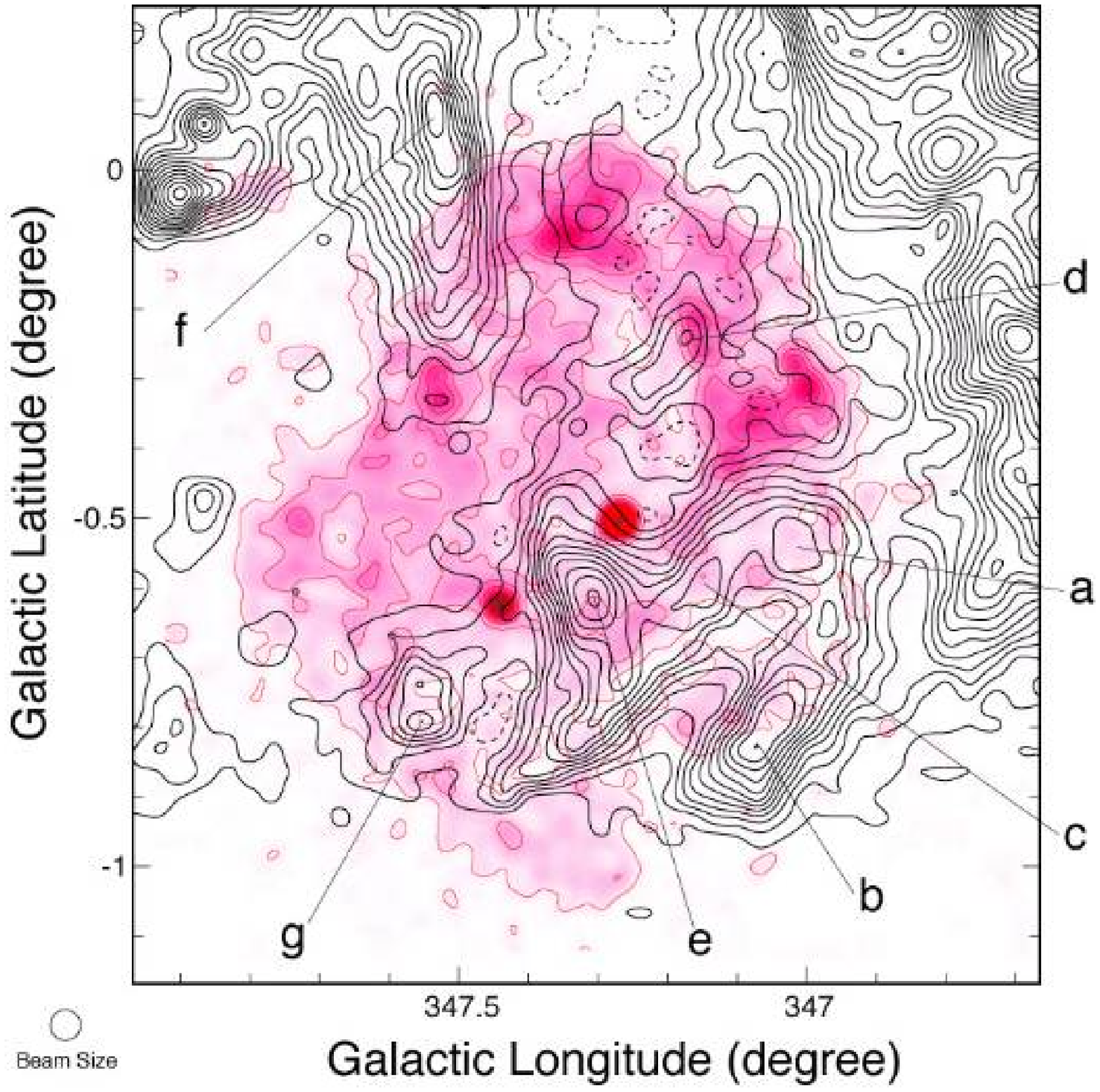}
\caption{A CO intensity map integrated in the velocity range from 6.5 km s$^{-1}$ to 7.5 km s$^{-1}$ overlapped with the ROSAT X-ray image same with Figure 2 in gray scale (colored pink in elctronic edition).  The positions of the local intensity peaks are denoted by (a)$-$(g).  The lowest contour level and interval are 0.7 K km s$^{-1}$ and 0.5 K km s$^{-1}$, respectively.  The depressions of the CO emission with the lowest contour level inside of the molecular boundary are shown with dashed contours to avoid confusion. \label{fig12}}
\end{figure*}


\clearpage


\begin{table}
\caption{Properties of $^{12}$CO($J=$1--0) Clouds}
\begin{tabular}{lccccccccr}
\hline \hline
Name & $l$ & $b$ & $T_{\rm R}^{*}$ & $V_{\rm peak}$ & ${\Delta}V_{\rm LSR}$ & Size & $L_{\rm co}$ & Mass & Comment\\
 & deg & deg & K & km s$^{-1}$ & km s$^{-1}$ & pc & $10^{2}$ K km s$^{-1}$ pc$^{2}$ & $M_{\odot}$& \\
\hline

A&346.933&-0.300&8.5&-10.3&4.8&4.2& 32.5 &686&peak A\\
B&346.933&-0.500&4.2&-8.0&4.6&2.1& 9.0 &190&peak B\\
C&347.067&-0.400&9.4&-12.0&3.8&3.0& 18.8 &397&peak C\\
D&347.300&0.000&4.0&-10.1&4.8&3.0& 13.9 &292&peak D\\
E&347.033&-0.200&2.0&-6.1&7.2&3.0& 7.5 &159&\\
F&346.867&-0.100&2.3&-3.5&5.0&2.7& 6.2 &131&\\
G&347.033&-0.067&3.3&-10.8&8.0&2.7& 14.6 &307&\\
H&346.933&0.033&1.5&-6.5/-8.5&-&2.4& 5.1 &109&\\
I&347.200&-0.233&1.8&-9.9&5.4&2.4& 4.9 &103&\\
J&347.300&0.133&5.5&-11.3&3.4&3.0& 15.5 &326&\\
K1&347.467&0.133&1.2&-4.3&3.6&2.7& 2.3 &48&\\
K2&347.467&0.133&2.2&-11.0&4.8&2.7 & 7.6 &160&\\
L&347.433&-0.133&4.0&-12.0&5.7&3.0& 17.6 &370&\\
M&347.867&-0.100&2.0&-9.5&6.3&3.3& 14.8 &312&\\
N1&347.767&-0.200&1.5&-4.0&4.0&2.4& 8.5 &178&\\
N2&347.767&-0.200&2.7&-10.9&6.1&2.4& 3.9 &83&\\
O&347.600&-0.333&1.1&-6.4&4.9&2.4& 2.9 &61&\\
P&347.933&-0.333&1.4&-8.3&9.0&2.7& 6.7 &141&\\
Q&347.800&-0.533&2.9&-2.8&3.2&3.0& 5.1 &108&\\
R&347.667&-0.733&4.1&-3.3&2.4&2.7& 3.2 &67&\\
S&347.400&-1.133&3.3&-4.8&1.5&1.8& 1.8 &38&\\
T&347.200&-0.967&3.3&-4.7&2.1&2.7& 2.9 &62&\\
U&347.167&-0.833&3.7&-4.8&1.3&2.4& 2.7 &58&\\
V&347.067&-0.933&3.6&-4.3&2.2&2.4& 2.7 &57&\\
W&346.933&-0.800&5.0&-5.1&3.0&4.0& 19.0 &402&\\
X&346.733&-0.867&5.7&-5.6&3.4&4.2& 27.8 &586&\\
Y&346.867&-0.667&3.4&-4.5&3.0&1.8& 4.2 &88&\\

\hline
\end{tabular}
\\
\\

Notes: Column (1): Cloud name. Column (2--3): Position of the observed point with the maximum $^{12}$CO($J$=1--0) intensity peak. Column (4--6): Observed properties of the $^{12}$CO($J$=1-0) spectra obtained at the peak positions of the CO clouds. Column (4): Peak radiation temperature $T_{\rm R}^{*}$. Column (5): $V_{\rm LSR}$ derived from a single Gaussian fitting.  Column (6): FWHM line-width ${\Delta}V_{\rm peak}$.  Column (7): Size defined as an effective diameter = (A/$\pi$)$^{0.5} \times$ 2, where A is the total cloud surface area defined as the region surrounded by the contour of 8.5 K or 6.5 K (See text).  If the contour is unclosed, the boundary is defined as the intensity minimum between the nearby peaks.  Column (8): The CO Luminosity of the cloud $L_{\rm co}$.  Column (9): Mass of the cloud derived by using the relationship between the molecular hydrogen column density $N(\rm H_{2})$ and the $^{12}$CO($J$=1--0) intensity $W(^{12}$CO), $N({\rm H_2}) = 2.0 \times 10^{20}\ [W({\rm ^{12}CO})/{\rm (K\ km\
s^{-1})}]\ {\rm (cm^{-2})}$ (Bertsch et al. 1993).  Column (10): The names of the peaks A$-$D identified in Paper I are noted.\\

\label{tab1}
\end{table}

\begin{table}
\caption{Results of LVG Analysis at Molecular Peaks}
\begin{tabular}{lccccccr}
\hline \hline
Name & $l$ & $b$ & $T_{{\rm CO}(J=1-0)}$ & $T_{{\rm CO}(J=3-2)}$ & $R(3-2/1-0)$ & $T_{\rm kin}(n=10^{3}/n=10^{4})$\\
 & deg & deg & K & K & & K \\
\hline

peak A&346.94&-0.32 & 8.9 & 7.5 & 0.8 & 50 / 30 \\
peak C&347.08&-0.40&9.3 & 6.5 & 0.7 & 40 / 25 \\
peak D&347.30&0.00&4.2 & 2.0 & 0.5 & 30 / 14\\

\hline
\end{tabular}
\\
\\

Notes: Column (1): Cloud name. Column (2--3): Position of the $^{12}$CO($J$=3--2) intensity peak. Column (4): Peak radiation temperature $T_{\rm R}^{*}$ of $^{12}$CO($J$=1--0) emission. Column (5): Peak radiation temperature $T_{\rm R}^{*}$ of $^{12}$CO($J$=3--2) emission averaged in the region within a contour at half-intensity levels for comparison with that of $J$=1--0.  Column (6): The ratio of $T_{\rm CO}(J$=3--2) to $T_{\rm CO}(J$=1--0) considering dilution effect of $J=$1--0 beam.  Column (7): Kinetic temperature assuming $n(\rm H_{2})$ = $10^{3}$ cm$^{-3}$ (left) and  $n(\rm H_{2})$ = $10^{4}$ cm$^{-3}$ (right).\\

\label{tab2}
\end{table}

\begin{table}
\caption{IRAS Point Sources toward Molecular Peaks}
\begin{tabular}{lccccccccr}
\hline \hline
IRAS name & $l$ & $b$ & $F_{12}$ & $F_{25}$ & $F_{60}$ & $F_{100}$ & $L_{\rm IRAS}$ & comment\\
& deg & deg & Jy & Jy & Jy & Jy & $L_{\odot}$&\\

\hline

17082--3955 &346.94&-0.31& 5.4 & 3.8 & 17.5 & 138 & 137 &peak A\\
17089--3951 &347.08&-0.39& 4.4 & 13.0 & 98.5 & 234 & 311 &peak C\\
17079--3926 &347.31&0.01& 2.0 & 20.0 & 88.6 & 739 & 562 &peak D\\

\hline
\end{tabular}
\\
\\

Notes: Column (1): IRAS source name. Column (2--3): Position of the source.  Column (4--7): Fluxes of 12, 25, 60, 100 $\mu$m band, respectively.  Column (8): IRAS luminosity estimated using formula of Emerson (1988).  Column (9): Nearby $^{12}$CO($J$=3--2) peak. (See text)\\

\label{tab3}
\end{table}

\begin{table}
\caption{$^{12}$CO($J$=1--0) Outflow Properties}
\begin{tabular}{lccr}
\hline\hline
&blue&red\\
\hline
Integrated intensity (K km s$^{-1}$) & 21.5 & 8.7\\
Mass ($M_{\odot}$) & 18 & 7\\
Size (arcmin) & 4.5 & 3.7\\
Size (pc) & 1.3 & 1.1\\
$\Delta V$ (km s$^{-1}$) & 9.0 & 9.0\\
$P$ ($M_{\odot}$ km s$^{-1}$) & 162 & 63\\
$E_{\rm kin}$ (10$^{45}$ erg) & 15 & 6\\
$t_{\rm dyn}$ (10$^{5}$ yr) & 1.5 & 1.2\\
$L_{\rm mech}$ ($L_{\odot}) $& 1.8 & 0.8\\
$F_{\rm co}$ (10$^{-3}$ $M_{\odot}$ km s$^{-1}$ yr$^{-1}$) & 1.1 & 0.5\\

\hline
\end{tabular}
\\
\\

Notes: The total intensity is estimated integrating over the area enclosed by a contour of 2.0 K for the integrated velocity range of $-$25 km s$^{-1}$ $\leq$ $V_{\rm LSR}$ $\leq$ $-$16 km s$^{-1}$ for the blue-shifted component, and $-$8 km s$^{-1}$ $\leq$ $V_{\rm LSR}$ $\leq$ $-$6 km s$^{-1}$ for the red-shifted one, respectively.  Size is defined as an effective diameter = ($A$/$\pi$)$^{0.5} \times$ 2, where $A$ is the region enclosed by a contour of 2.0 K.  Mass is derived by using the relationship between the molecular hydrogen column density $N(\rm H_{2})$ and the $^{12}$CO($J$=1--0) intensity $W(^{12}$CO), $N({\rm H_2}) = 2.0 \times 10^{20}\ [W({\rm ^{12}CO})/{\rm (K\ km\ s^{-1})}]\ {\rm (cm^{-2})}$ (Bertsch et al. 1993).  $\Delta V$ is a velocity range of the wing component estimated from blue-shifted lobe (for red-shifted, the same value with the blue-shited one is adopted tentatively).  $P$ (momentum), $E_{\rm kin}$ (kinematic energy), $t_{\rm dyn}$ (dynamical age), $L_{\rm mech}$ (mechanical luminosity), and $F_{\rm co}$ (CO momentum flux) is defined as Mass $\times$ $\Delta V$, 1/2$M\Delta V^{2}$, Size/$\Delta V$, 1/2$M\Delta V^{3}/R$, and $P$/$t_{\rm dyn}$, respectively.
\\

\label{tab4}
\end{table}

\begin{table}
\caption{Properties of Local Cloud Peaks}
\begin{tabular}{lcccccccr}
\hline\hline
Name & $l$ & $b$ & $V_{\rm LSR}$ & $T_{\rm R}^{*}$ & $\Delta V$ & I.I. & $n({\rm H})_{\rm CO}$\\ 

& deg & deg & km s$^{-1}$ & K & km s$^{-1}$ & K km s$^{-1}$ & 10$^{21}$ cm$^{-2}$ \\ 

\hline

a&347.02&-0.54&6.4&9.4&1.3&11.8&4.7\\ 
b&347.07&-0.83&7.0&5.7&1.2&6.9&2.8&\\ 
c&347.15&-0.60&6.7&6.7&1.5&10.1&4.0\\ 
d&347.17&-0.24&7.1&2.9&2.1&6.0&2.4\\ 
e&347.30&-0.62&7.0&7.3&1.7&12.6&5.0\\ 
f&347.53&0.08&7.1&7.7&1.6&11.9&4.8\\ 
g&347.55&-0.79&6.3&6.9&1.1&7.7&3.1\\ 

\hline
\end{tabular}
\\
\\

Notes: Column (1): Peak name. Column (2--3): Position of the $^{12}$CO($J$=1--0) peak.  Column (4): $V_{\rm LSR}$ derived from a single Gaussian fitting.  Column (5): Peak radiation temperature $T_{\rm R}^{*}$.  Column (6): FWHM line-width ${\Delta}V_{\rm peak}$.  Column (7): $^{12}$CO($J$=1--0) integrated intensity from a single Gaussian fitting.  Column (8): Hydrogen column density estimated from CO intrensity assuming an X-factor of 2 $\times 10^{20}$ cm$^{-2}$ K$^{-1}$ (km s$^{-1}$)$^{-1}$ Bertsch et al. (1993). \\

\label{tab5}
\end{table}

\begin{table}
\caption{Physical Poperties of G347.3--0.5}
\begin{tabular}{l@{  ...  }cc}
\hline \hline
Parameters & 1 kpc 	& 6 kpc 	\\
\hline
Diameter (60') 		& 17.4 pc 	& 104 pc 	\\
Histrical record 	& A.D. 393$^{*1}$ 	& -- 		\\
Age (year)		& 1,600 	& $>$10,000 	\\
Evolution phase 	& Free-expansion & Sedov 	\\
Ambient density (cm$^{-3}$) & $<$ 0.01 	& 0.003$^{*2}$ 	\\
Shock velocity (km s$^{-1}$) & 5,500 	& 3,200$^{*2}$	\\
Shock temperature (keV) & 35 		& 12$^{*2}$ 	\\
Ejecta			& Non-radiative	& Radiative 	\\
Swept-up mass ($M_{\odot}$) & $<$ 3 	& 35$^{*2}$ 	\\  
E.I. (cm$^{-5}$) 	& $<5 \times 10^{16}$ & $2 \times 10^{15}\;^{*2}$\\ 
Central source mitting area (km)	& 3 		& 0.5 \\
Total energy of accelerated particle (erg) & $\sim 2 \times 10^{49}$ & $\sim
10^{50}$ \\  
\hline
\end{tabular}
{
\\ 1. Wang, et al. 1997.\\
2. Assuming 2 $\times$ 10$^{4}$ years of age.
The physical parameters for 1 kpc and 6 kpc distance are
shown in this table. The remnant age for 6 kpc is assumed 20000 year
to estimate ambient density, shock velocity, shock temperature,
swept-up mass and emission measure.}

\label{tab6}
\end{table}

\end{document}